\documentclass[11pt,a4paper]{article}
\usepackage{jcappub}

\usepackage{graphicx}	
\usepackage{amsmath}	
\usepackage{bm}
\usepackage{color}
\usepackage{CJKutf8}

\newcommand{\be}{\begin{equation}}
\newcommand{\ee}{\end{equation}}
\newcommand{\bea}{\begin{eqnarray}}
\newcommand{\eea}{\end{eqnarray}}
\def\ba#1\ea{\begin{align}#1\end{align}}

\newcommand{\refeq}[1]{Equation (\ref{eq:#1})}
\newcommand{\refeqs}[2]{Equations(\ref{eq:#1})--(\ref{eq:#2})}
\newcommand{\reffig}[1]{Figure \ref{fig:#1}}

\newcommand{\refsec}[1]{Section~\ref{sec:#1}}
\newcommand{\refapp}[1]{Appendix ~\ref{app:#1}}

\newcommand{\vs}{\nonumber\\}

\def\({\left(}
\def\){\right)}
\def\[{\left[}
\def\]{\right]}
\def\<{\left<}
\def\>{\right>}

\def\vr{\bm{r}}
\def\vx{\bm{x}}

\def\vvs{\bm{s}}
\def\vk{{\bm{k}}}
\def\vq{{\bm{q}}}

\def\khat{\hat{\bm{k}}}

\def\d{{\rm d}}

\def\max{{\rm max}}

\def\Plin{P_{L}}

\def\bPsi{\boldsymbol{\Psi}}

\definecolor{RedWine}{rgb}{0.743,0,0}
\definecolor{GrassGreen}{rgb}{0.125,0.75,0.125}
\definecolor{RoyalBlue}{rgb}{0.25,0.41,0.88}
\definecolor{NavyBlue}{rgb}{0,0,0.9}

\definecolor{RedWine}{rgb}{0.743,0,0}

\definecolor{NavyBlue}{rgb}{0,0,0.9}

\def\apjl{Astrophys.\ J.\ Lett.}

\def\aap{Astron.\ Astrophys.}

\def\apjs{Astrophys.\ J. Supp.}

\title%
{Reconstructing the Long-wavelength Matter Density Fluctuation Modes from the Scalar-Type Clustering Fossils}

\author[1,2]{Zhenyuan Wang (\begin{CJK*}{UTF8}{gbsn}王震远\end{CJK*})}
\author[1,2,3]{Donghui Jeong}

\affiliation[1]{Department of Astronomy and Astrophysics, The Pennsylvania State University, University Park, PA 16802, USA}
\affiliation[2]{Institute for Gravitation and the Cosmos, The Pennsylvania State University, University Park, PA 16802, USA}
\affiliation[3]{School of Physics, Korea Institute for Advanced Study, 85 Hoegiro, Dongdaemun-gu, Seoul, 02455, Republic of Korea}

\emailAdd{zzw173@psu.edu}
\emailAdd{djeong@psu.edu}

\abstract{
Revealing the large-scale structure from the 21cm intensity mapping surveys is only possible after the foreground cleaning. However, most current cleaning techniques relying on the smoothness of the foreground spectrum lead to a severe side effect of removing the large-scale structure signal along the line of sight. On the other hand, the clustering fossil, a coherent variation of the small-scale clustering over large scales, allows us to recover the long-wavelength density modes from the off-diagonal correlation between short-wavelength modes. In this paper, we revisit the reconstruction based on the short-wavelength matter density modes in real space and scrutinize the requirements for an unbiased and optimal clustering-fossil estimator. We show that (A) the estimator is unbiased only when using an accurate bispectrum model for the long-short-short mode coupling and (B) including the connected four-point correlation functions is essential for characterizing the noise power spectrum of the estimated long mode. For matter in real space, the clustering fossil estimator based upon the leading-order bispectrum  yields an unbiased estimation of the long-wavelength ($k\lesssim 0.01~[h/{\rm Mpc}]$) modes with the cross-correlation coefficient of $0.7$ at redshifts $z=0$ to $3$.
}


\arxivnumber{2312.17321}

\begin{document}
\maketitle

\section{Introduction}\label{sec:intro}
The intensity mapping of the 21cm hyper-fine transition line (21cmIM) can be a powerful probe of the large-scale structure of the Universe \citep{chang2008baryon, loeb2008possibility}. With the 21cm line's low opacity and spectroscopic nature, the 21cmIM will provide a three-dimensional map of neutral hydrogen distribution over the unprecedently large volume. The ongoing and future 21cmIM projects, such as 
Tianlai\citep{chen2012Tianlai}, 
CHIME\citep{bandura2014CHIME}, 
HIRAX\citep{newburgh2016hirax}, 
BINGO\citep{battye2013ra}, 
and SKA\citep{Pritchard2015}, are designed to observe the large-scale structure at redshifts $z = 0.5 - 6$, to measure, for example, the Hubble expansion rate and the angular diameter distance from the baryon acoustic oscillation (BAO) \citep{seo2010ground}.

One of the main challenges of these 21cmIM experiments is the contamination from the extremely loud foreground from synchrotron, free-free, and thermal dust emissions. For example, the dominating galactic synchrotron emission is more than five orders of magnitude louder compared to the targeted large-scale structure signal in 21cmIM \citep{chapman2019, liu2020data}. The standard foreground cleaning method \citep{Morales2006, Wang2006, Bowman2009, Liu2009, Liu2011, Parsons2012, Chapman2012, Chapman2013, Dillon2013, Wolz2017, Carucci2020} takes advantage of the fact that synchrotron radiation is smooth in the frequency domain, while the 21cmIM signals show genuine cosmological density fluctuations along the radial direction. After the foreground cleaning, however, the long-wavelength Fourier modes parallel to the line of sight become inaccessible \citep{Furlanetto2006, liu2020data}. Furthermore, the frequency-dependence response of the instrument can lead to a {\it foreground wedge} contamination in the Fourier space for $k_\parallel<k_\perp \tan\psi$, where the angle $\psi$ increases with the field of view \citep{morales2012four, ade2016planck, Trott2012, Dillon2014, Liu2014, Pober2014, Pober2015}.
 
These long-wavelength modes plagued by foreground cleaning are usually in the linear regime where the observed galaxy clustering statistical can be modeled by the Kaiser formula \cite{Kaiser:1984,Kaiser:1987}, or near horizon scales where general relativistic corrections are important (see Ref.~\cite{Jeong/Schmidt:2015} for a review). For both cases, the theoretical model is well understood in the framework of linear perturbation theory, and the accurate measurement of these long-wavelength modes will be translated to the measurement of the growth rate of the cosmic density field and metric perturbations as well as the local-type primordial non-Gaussianities \cite{Desjacques2018}. Measuring these parameters, in turn, will inform us of the nature of dark energy and the physics of the primordial universe, respectively. In particular, given the large volume that 21cmIM covers, the large number of long-wavelength modes could provide an unprecedented tight constraint on cosmological parameters \cite{McQuinn2006,Mao2008,Bull2015,Karagiannis2020}. Therefore, to fully exploit the information from 21cmIM, it is crucial to recover the contaminated long-wavelength Fourier modes (hereafter, ``long mode,'' for short).

To recover the contaminated long modes, Ref.~\cite{pen2012cosmic} have proposed an innovative technique called {\it cosmic-tide} reconstruction. The basic idea of the reconstruction is to exploit the non-Gaussianity of the density field coming from the nonlinear gravitational evolution that couples Fourier modes of different wavelengths. Namely, knowing the details of this coupling should allow us to infer the long-wavelength Fourier modes from the short-wavelength Fourier modes (hereafter,  ``short mode,'' for short). Based on this idea, Ref.~\cite{Zhu_2016} developes a quadratic estimator for the long modes based on the anisotropic modulation of short modes' power spectrum by the large-scale tidal field. The coupling coefficient in the estimator is determined by the leading-order tidal interaction \citep{schmidt2014tidal}. Applying this quadratic estimator to the $N$-body simulation results shows that the cross-correlation coefficient between the reconstructed long mode and the original long mode can reach 0.9 at $k<0.1$ $h/$Mpc. This implies that the phase of the reconstructed long mode is highly correlated to the true long mode. Ref.~\cite{Karacayli_2019} has further confirmed this result. Additional insights into tide reconstruction from tracers in real and redshift space are provided by \cite{Zhu2022} and \cite{Zang2022}. Also see \citep{Zhu2018} for its application to recovering missing radial long modes in 21cm intensity mapping surveys, and \citep{Foreman2018} in the context of lensing reconstruction of line intensity mapping.

While following the spirit of cosmic-tide reconstruction, this paper investigates a different mathematical framework, called {\it clustering fossil}, for reconstructing the long modes. The clustering fossil utilizes the three-point non-Gaussian correlation between long mode and two short modes, which generates, in Fourier space, the off-diagonal two-point correlators between two short modes \citep{jeong2012clustering}. In a statistically homogeneous universe, the two-point correlation function $\left<\delta^*(\vk)\delta(\vk')\right>$ in Fourier space only has diagonal components proportional to $\delta^D(\vk-\vk')$: the power spectrum. In the presence of long modes, however, the non-Gaussian coupling between short modes and long modes makes locally measured two-point correlators differ from place to place. The statistical homogeneity is broken up locally, and the non-zero off-diagonal two-point correlators among short modes contain information about the long mode coupling with them. As we show in \refsec{fossil}, this off-diagonal correlation is proportional to the amplitude of the long mode and the coupling coefficient, which can be read from the squeezed-limit bispectrum.

Following Ref.~\citep{jeong2012clustering}, we construct a quadratic clustering-fossil estimator for the long modes based on the off-diagonal correlation between short modes. Unlike Ref.~\citep{jeong2012clustering}, which pursues inflationary spectator fields using the non-Gaussian coupling in the early universe \cite{dai2013seeking,dimastrogiovanni2014inflationary}; however, we focus on the non-Gaussianities caused by nonlinear matter clustering. In particular,  applying the method to the case of 21cmIM, we treat the contaminated long mode as a scalar-type fossil field. Along this line, Refs.~\cite{Li2020reconstruction_a, Li2020new} have applied the clustering-fossil estimator to reconstruct the long mode based on the leading-order (tree-level) bispectrum in standard cosmological perturbation theory (SPT) \citep{bernardeau2002large}. Analyzing a suite of $N$-body simulations, they have recovered the long mode, with morphological features similar to the ground truth. They found, however, that the power spectrum of the recovered long mode is biased, which is consistent with the findings of Ref.~\cite{Zhu_2016}. A further study \citep{darwish2021density} has applied this technique to the biased tracer fields to recover the matter density long mode and forecast the improved constraint on primordial non-Gaussianity through multi-tracer method, where the galaxy bias are also included into the estimator.

To proceed with the clustering-fossil-based reconstruction method, we need a more thorough theoretical understanding of the method's applicability and regime of validity. The goal of this paper is to scrutinize the method. We aim to provide a theoretical framework to explain the systematic bias observed in the N-body simulations and to characterize the statistical uncertainties of the recovered long mode. To solve the problem for the fully nonlinear density field, we need to include the coupling effects between long and short modes beyond the leading order and the non-Gaussian statistics of the short modes. To assess the requirement for a robust reconstruction method, we take a simpler approach here. Namely, instead of analyzing the result from full $N$-body simulations, we test the reconstruction method against a {\it controlled sample} density field only containing the first- and second-order density perturbations from the GridSPT (grid-based calculation of standard perturbation theory) \citep{Taruya2018Grid, Taruya2021Grid, Taruya2022GridRSD}. Because the GridSPT density field strictly follows SPT, the fossil estimator constructed from the leading-order bispectrum can fully capture the coupling between the long and short modes. Therefore, testing the clustering fossil estimator in this way allows us to disentangle the effect of higher-order coupling from other effects, such as the non-Gaussianity of short modes. To inject the effect from the higher-order nonlinear couplings, we use the 2LPT (second-order Lagrangian Perturbation Theory) because, while agreeing with the SPT to second order, the 2LPT density field also contains higher-order contributions. We take advantage of the efficiency of both GridSPT and 2LPT, which enable us to generate a large number of realizations to suppress the sampling variance and find any underlying systematic errors in the estimator. 

By testing the estimator on well-controlled datasets, we find that the clustering-fossil estimator is unbiased only when an accurate bispectrum model is used for the coupling between long and short modes. Furthermore, to reconstruct the long-mode power spectrum, the connected four-point correlation function must be included to compute the noise power spectrum. It is important to note that the goal of this paper is to assess the potential systematics of the quadratic estimator, so we restrict our study to matter clustering in real space for clarity and transparency. To apply this method to galaxy surveys or 21cm intensity mapping experiments, we need to include the galaxy bias and redshift space distortions into account(see e.g. \cite{darwish2021density}, \cite{Zang2022}), which would make the accurate reconstruction even harder.

The rest of this paper is organized as follows. In \refsec{method}, we review the clustering fossil technique and construct the optimal estimator of the large-scale matter density modes. In \refsec{implement}, we introduce GridSPT and 2LPT we use to generate the nonlinear short-wavelength modes and implement the fossil estimator to reconstruct the large-scale mode. We compare the results with the theoretical prediction in \refsec{results}. We conclude and discuss the future applications in \refsec{conclusion}. We provide the fossil estimator in the continuous limit in \refapp{fossil_continuous}.

Throughout this paper, we use the following convention of Fourier transformation,
\bea
&&f(\vk) = \int \d^3 x f(\vx) e^{-i\vk\cdot \vx},
\\
&&f(\vx) = \int \frac{\d^3 k}{(2\pi)^3} f(\vk) e^{i\vk\cdot \vx}.
\eea
Note that we use the same character for a function $f$ and distinguish the Fourier-space representation and configuration-space representation by the argument. For the compactness of the equations, we use the following convention for the sum of multiple vectors,
\bea
\vk_{1\cdots n} = \sum_{i=1}^n \vk_i.
\eea

\section{Estimating long modes from the clustering-fossil estimator}
\label{sec:method}
We begin the section by motivating the clustering-fossil method \citep{jeong2012clustering} for a generic non-Gaussian density field in \refsec{general_fossil}. In \refsec{fossil}, we construct the optimal quadratic estimator for measuring the long mode from the imprint of the squeeze-limit bispectrum, or long-short-short coupling. We also present the power spectrum of the reconstructed long modes and the cross-correlation coefficient between the original and the reconstructed long modes. Finally, we present the reconstructed long modes' covariance matrix, or noise power spectrum, in \refsec{fullPN}.

\subsection{Position-dependent correlations induced by squeezed-limit non-Gaussianity}
\label{sec:general_fossil}

The standard cosmological model based on the Friedman-Lema\^itre-Robertson-Walker world model is spatially homogeneous and isotropic. The spatial homogeneity extends to the clustering properties of galaxies, which is often referred to as statistical homogeneity. We begin this section by defining the statistical homogeneity through $n$-point correlation functions.

The fundamental quantity describing the galaxy clustering is the density contrast $\delta(\vx)\equiv n(\vx)/\bar{n}-1$, where $n(\vx)$ is the density at location $\vx$ and $\bar{n}$ is the cosmic mean density. The two-point correlation function is defined in terms of the density contrast as
\bea
\xi(\vr) = \< \delta(\vx) \delta(\vx + \vr)\>\,.
\eea
Here, the bracket $\<...\>$ represents the average over the statistical ensemble of the cosmic density field. Note that $\xi(\vr)$ depends only on the separation vector $\vr$ between two positions $\vx$ and $\vx+\vr$ but is independent of the position $\vx$ where we make the measurement. This property manifests the statistical homogeneity. Furthermore, the statistical isotropy would restrict $\xi(\vr)=\xi(r)$. Equivalently, the power spectrum, the Fourier transformation of the two-point correlation function, is defined as
\bea
\< \delta(\vk_1) \delta(\vk_2)\> = (2\pi)^3 \delta^D(\vk_{12}) P(\vk_1),
\eea
where the Dirac-delta operator, $\delta^D(\vk_{12})$ on the right-hand side, encodes the statistical homogeneity. That is, the statistical homogeneity dictates that the correlation between the Fourier modes $\delta(\vk_1)$ and $\delta(\vk_2)$ vanishes unless $\vk_1 + \vk_2 = 0$. We call such two-point correlators in Fourier space {\it diagonal}.

We can extend this concept and write the $n$-point correlation function and the $n$-poly spectra for statistically homogeneous density contrast $\delta$ as follows:
\bea
&&\zeta^{(n)}(\vr,\vvs,\cdots)
=
\<
\delta(\vx)\delta(\vx+\vr)\delta(\vx+\vvs)\cdots
\>
\label{eq:defxin}
\\
&&\<
\delta(\vk_1)\cdots\delta(\vk_n)
\>
=
(2\pi)^3 \delta^D(\vk_{12\cdots n}) 
P^{(n)}(\vk_1,\cdots,\vk_n)\,.
\label{eq:defpkn}
\eea

We call a density field Gaussian when the two-point correlation function is the only non-vanishing connected $n$-point correlation function \cite{Jeong:thesis}; otherwise, the density field is non-Gaussian. 

The clustering-fossil estimator exploits the fact that non-Gaussianities in density fields can yield a spatial variation of correlation functions. That is, even if the density field satisfies statistical homogeneity when taking the ensemble average as in \refeq{defxin}, the locally measured $n$-point correlation function can have an explicit position dependence through the non-Gaussian correlation. This is because the non-Gaussian correlation function involving one or more long-wavelength modes ($\delta_\ell$) can be broken down as the multiplication of the conditional correlation functions and the conditioning long modes:
\bea
&&\left\langle
\delta_\ell(\vx_1)\cdots\delta_\ell(\vx_n)\delta(\vr)\delta(\vvs)\cdots
\right\rangle
=
\left\langle
\left.\left\langle\delta(\vr)\delta(\vvs)\cdots
\right\rangle
\right|_{\delta_\ell(\vx_1')\cdots\delta_\ell(\vx_n')}
\delta_\ell(\vx_1)\cdots\delta_\ell(\vx_n)
\right\rangle\,,
\eea
where $\langle\rangle|_{X}$ stands for the correlator evaluated with the condition $X$. The expression follows from the definition of conditional probability $P(A\cap B) = P(A|B)P(B)$.

The most well-studied example of clustering fossil comes from the squeezed (or soft) limit of non-Gaussian correlation functions. Namely, the squeezed-limit $(n+1)$-point correlation function can modulate the $n$-point correlation function, inducing the position-dependent power spectrum \citep{chiang2014position, chiang2015position} or the position-dependent bispectrum \citep{adhikari2016constraining}. In this case, we may write the conditional correlation function as
\bea
&&
\left.\left\langle\delta(\vr)\delta(\vvs)\cdots
\right\rangle
\right|_{\delta_\ell(\vx')}
=
C
+
\left.
\frac{\d}{\d \delta_\ell}
\left\langle\delta(\vr)\delta(\vvs)\cdots
\right\rangle
\right|_{\delta_\ell=0}
\delta_\ell(\vx')
+
\cdots\,,
\label{eq:n-ptfn}
\eea
where $C$ is a function independent from $\delta_\ell$, and we truncate the expansion at the linear-order response. On large scales where the long mode $\delta_\ell$ is in the linear regime, we compute the original correlation function as
\bea
&&
\left\langle
\delta_\ell(\vx)\delta(\vr)\delta(\vvs)\cdots
\right\rangle
\simeq
\left.
\frac{\d}{\d \delta_\ell}
\left\langle\delta(\vr)\delta(\vvs)\cdots
\right\rangle
\right|_{\delta_\ell=0}
\xi_\ell(\vx-\vx')\,.
\label{eq:n+1-ptfn}
\eea
The general idea of the clustering fossil estimator \citep{jeong2012clustering} is to use \refeq{n-ptfn} as a starting point for estimating the long-wavelength mode $\delta_\ell$ for the non-Gaussian density field with underlying $(n+1)$-point correlation function taking the form of \refeq{n+1-ptfn}. 

As shown in \citep{jeong2012clustering}, the same argument applies when using a long mode $h_\ell^s$ with general spin $s$. The density field is a spin-0 case.

\subsection{The Optimal Clustering Fossil estimator}
\label{sec:fossil}
Let us focus on the quadratic clustering-fossil estimator using the squeezed-limit three-point correlation function, or its Fourier transform, bispectrum. The Fourier-space counterpart of \refeq{n+1-ptfn}, when considering the three-point correlator, becomes
\bea
\label{eq:squeezedbk}
\<\delta_\ell(\vk)\delta(\vk_1)\delta(\vk_2)\> 
=
(2\pi)^3 \delta^D(\vk_{12} + \vk) f(\vk_1, \vk_2; \vk) P_\ell(k),
\label{eq:squeezedBk}
\eea
with kernel $f( \vk_1 , \vk_2; \vk)$ encoding the details of the coupling between the long mode $\delta_\ell(\vk)$ and two short modes $\delta(\vk_1)$ and $\delta(\vk_2)$. Hereafter, we drop the explicit subscript $\ell$ and implicitly use $\vk\ll\vk_i$ to indicate the long mode. We emphasize that the three wave vectors in $f(\vk_1 , \vk_2; \vk)$ must satisfy $\vk_{12} + \vk = 0$.

As we have shown in \refsec{general_fossil}, \refeq{squeezedBk} is equivalent to the following conditional two-point correlator:
\bea
\label{eq:fossileq}
\<\delta(\vk_1)\delta(\vk_2)\>\big|_{\delta(\vk)} = f(\vk_1, \vk_2;\vk) \delta^*(\vk)\delta^{\rm K}_{\vk_{12}, -\vk}.
\eea
Here, $\delta^{\rm K}$ refers to the Kronecker delta that we use instead of Dirac delta, after absorbing appropriate dimensionful constants inside $f(\vk_1,\vk_2;\vk)$, and the superscript $*$ is the complex conjugate. The locally broken statistical homogeneity is now manifested by the off-diagonal correlator in \refeq{fossileq}.

Note that \refeq{squeezedbk}, more generally \refeq{n+1-ptfn}, assumes that the long mode is in the linear regime where we can neglect the higher order terms in \refeq{n+1-ptfn}; so the fossil estimator is unbiased only in that limit. A generic unbiased fossil estimator would require additional corrections to include higher-order effects. In \refsec{limit} of this paper, however, we show that the nonlinear part of the long-mode power spectrum is less than a few percent of the cosmic variance, which ensures that the nonlinear correction is unimportant at the scale within with we implement the estimator.

Let us translate the fossil equation, \refeq{fossileq}, to the quadratic estimator for the long mode $\delta(\vk)$. A simple estimator turning \refeq{fossileq} around would be 
\bea
\hat\delta_r^{\rm naive}(\vk) 
= 
\left[
\sum_{-\vq_i} \frac{\delta(-\vq_i)\delta(-\vk+\vq_i)}{f(-\vq_i,-\vk+\vq_i;\vk)}
\right]^*
=
\sum_{\vq_i} \frac{\delta(\vq_i)\delta(\vk-\vq_i)}{f(\vq_i,\vk-\vq_i;-\vk)}\,.
\label{eq:naive}
\eea
In the second equality, we use the reality of the density field $\[\delta(\vk)\]^* = \delta(-\vk)$ and that bispectra and $f(\vk_1,\vk_2;\vk)$ must be real for even-parity density field: $f^*(\vk_1,\vk_2;\vk) = f(-\vk_1,-\vk_2;-\vk)$ \citep{Jeong/Schmidt:2020}. 
For a given $\vk$, this naive estimator sums over all possible pairs $\delta(\vq_i)$ and $\delta(\vk-\vq_i)$ weighted by a known kernel function $f(\vq_i,\vk-\vq_i;-\vk)$ determined by the bispectrum as in \refeq{squeezedbk}. In order to exclude the duplication of counting the same quadratic contribution twice, we demand the condition 
$k < |\vk - \vq_i| \le q_i$.
Since the estimator is based on the multiplication of two short modes, it is called a quadratic estimator. A similar quadratic estimator has already been used widely in the CMB lensing reconstruction \citep{hu2002mass, okamoto2003cosmic}.

Noticing that the individual contribution in \refeq{naive} is subject to different variances, we can introduce a normalized weight $W_i$ 
\bea
\hat{\delta}_r(\vk) 
=
\sum_i W_i \frac{\delta(\vq_i)\delta(\vk-\vq_i)}{f(\vq_i,\vk-\vq_i;-\vk)},
\label{eq:Wieq}
\eea
to find an optimal estimator which minimizes the variance of $\hat\delta_r(\vk)$:
\bea
\quad
\sigma^{2}\[\hat{\delta}_r(\vk)\]
=
\sum_{i j} W_{i} W_{j} C_{i j}.
\eea
First, we compute the covariance matrix $C_{ij}$ for individual contribution in the naive estimator [\refeq{naive}]:
\bea
C_{ij}(\vq_i, \vq_j;\vk) 
=
\label{eq:Cov_diag}
V^2 \frac{P(q_i)P(|\vk-\vq_i|)}{f^2(\vq_i,\vk-\vq_i;-\vk)} \delta^{\rm K}_{\vq_j, \,-\vq_i} ,
\eea
by ignoring all the connected parts of four-point correlators. Here $P(q)$ and $P(|\vk - \vq|)$ are the measured nonlinear power spectra of the short modes. In this case, the inverse-variance weight
\bea
\label{eq:weight_cov}
W_i = \sum_{j} C_{i j}^{-1} \big/ \sum_{i j} C_{i j}^{-1},
\;
\sigma^{2}\[\hat{\delta}_r(\vk)\]
= 1 \big/ \sum_{j} C_{i j}^{-1}.
\eea
gives the minimum variance estimator. Using the inverse-variance weight in \refeq{weight_cov}, the optimal fossil estimator for the long mode becomes
\bea
\label{eq:estimator}
\hat{\delta}_r (\vk) 
=
P^N_{G} (k) \sum_{\vq} \frac { f(\vq, \vk- \vq;-\vk) } { VP( \vq) P ( \vk- \vq) } \delta ( \vq) \delta (\vk- \vq),
\eea
the variance of the optimal estimator $\hat{\delta}_r$ becomes
\bea
\label{eq:PN_G}
P^N_{G}(k)
= 
\[ \sum_{\vq} \frac { \left| f ( \vq , \vk - \vq ;-\vk) \right| ^ { 2 } } { V P ( \vq ) P ( \vk - \vq ) } \]^{-1}\,.
\eea

The minimum variance $P^N_{G}(k)$ turns out to be the noise term of the recovered long mode's auto power spectrum, when taking only the Gaussian (disconnected) four-point correlator of the density contrast $\delta$:
\bea
{P}_{rr}(k)
\equiv 
\frac{1}{V}\frac{1}{N_k}\sum_{\khat} \< |\hat\delta_r(\vk)|^2 \> 
 = P_L(k) + P^N_G(k)\,.
 \label{eq:Prrk}
\eea
Here, $N_k$ is the number of Fourier modes for a fixed $k$. \refeq{Prrk} suggests that we have to subtract the noise power spectrum $P_G^N(k)$ from the auto power spectrum, in order to recover an unbiased long-mode power spectrum.

Let us turn into the cross-correlation coefficient $r(k)$ between the recovered long mode $\hat{\delta}_r(\vk)$ and the original one $\delta(\vk)$ that we designate with subscript $m$:
\bea
\label{eq:rk}
r(k) \equiv \frac{P_{rm}(k)}{\sqrt{{P}_{rr}(k)P_{mm}(k)}} 
\eea
Here, $\left<\hat{\delta}_r(\vk)\delta_m(\vk')\right> = (2\pi)^3 P_{rm}(k)\delta^D(\vk+\vk')$ is the cross power spectrum 
\bea
P_{rm}(k) 
\equiv
\frac1V \frac{1}{N_k}
\sum_{\khat} {\rm Re}\< \hat{\delta}_r(\vk) \delta^*(\vk) \> 
= \sum_{\vq_i} W_i \frac{B(\vq_i,\vk-\vq_i,-\vk)}{f(\vq_i,\vk-\vq_i;-\vk)}
\simeq P_L(k)\,,
\label{eq:Prm_leading}
\eea
which approaches the linear power spectrum for the squeezed ($k\to0$) limit. In \refsec{spt2nd}, we calculate the correction term to this approximation when using the full bispectrum expression.  The leading contribution of the correction term comes from the correlation between the second-order of the long mode and two first-order short modes, which vanishes in the squeezed limit.

Combining all results, we calculate the leading-order expression for the cross-correlation coefficients in the squeezed limit [$k\to0$ and $P_{mm}\to P_L(k)$] as,
\bea
\label{eq:Rk}
r(k) \simeq \frac{1}{\sqrt{1 + P^N_G(k)/P_L(k)}}.
\eea
From \refeq{Rk}, we see that $r(k)\to1$ if and only if when the noise power spectrum is much suppressed compared to the linear power spectrum. The noise power spectrum [\refeq{PN_G}] becomes smaller as we include more and more short modes in the quadratic estimator. Meanwhile, the noise power spectrum is insensitive to the lower limit of $q$ and $|\vk-\vq|$ because most reconstruction power comes from the high-frequency ($|\vq|\gg k$) modes. In this paper, we use the short modes $\delta(\vq),\, \delta(\vk-\vq)$ with wavenumbers satisfying
\bea
\label{eq:qrange}
k < |\vk-\vq| \le q \le q_\max,
\eea
where $q_{\rm max}$ is the wavenumber of the smallest scale used for reconstruction. Ideally, we want to make $q_{\rm max}$ as large as possible to suppress the noise $P^N_G$. In practice, however, we are limited by strong nonlinearities on small scales, so we can only keep $q_{\rm max}$ in the quasi-linear scale where perturbation theory accurately models the nonlinearities \citep{jeong2006perturbation,tomlinson2022Bk}. As we shall show below, using the short modes beyond the quasi-linear regime leads to the systematic bias in the reconstructed long modes.

As stated earlier, the summations in the equations above excludes the duplications (that is, $\vq'$=$\vk-\vq$ is identical to $\vq$), which leads to a factor of two difference between \refeq{PN_G} here and the expressions shown in previous studies \citep{jeong2012clustering, Li2020reconstruction_a,Li2020new, darwish2021density}. The restriction in the summation, however, cancels the factor-of-two difference, so the expressions are identical.

However, just like our derivation in this section, most previous studies to date have ignored the connected four-point correlation function in deriving \refeq{Cov_diag}, so the noise power spectrum is underestimated (see \reffig{PNz0z1} below). We now present the full noise power spectrum, including the connected four-point function. Readers with interest can find related discussions of connected four-point correlation function in the noise power spectrum of quadratic estimators in \citep{Foreman2018, darwish2021density} for long-mode reconstruction and \citep{Schaan2018} for CIB lensing.

\subsection{The Full Noise Power Spectrum}
\label{sec:fullPN}
In \refsec{fossil}, we use the Gaussian approximation for when calculating the covariance matrix of naive quadratic terms estimators in \refeq{naive}. In this section, we supplement that calculation by deriving the full covariance matrix, including the connected four-point function. We also present the improved noise power spectrum by using the full covariance matrix but still adopting the inverse-variant weight using the Gaussian approximation. 

We first compute the full covariance matrix of each term contributing to the naive estimator \refeq{naive} as
\begin{eqnarray}
C_{ij}(\vq_i, \vq_j;\vk) 
&=&
\<\frac{\delta(\vq_i)\delta(\vk-\vq_i)}{f(\vq_i, \vk-\vq_i;-\vk)} \frac{\delta(\vq_j)\delta(-\vk-\vq_j)}{f(\vq_j, -\vk-\vq_j;\vk)}\>\bigg|_{\delta(\vk),{\delta(-\vk)}}
\nonumber\\
&& - 
\<\frac{\delta(\vq_i)\delta(\vk-\vq_i)}{f(\vq_i, \vk-\vq_i;-\vk)}\>\bigg|_{\delta(\vk)}
\< \frac{\delta(\vq_j)\delta(-\vk-\vq_j)}{f(\vq_j, -\vk-\vq_j;\vk)}\>\bigg|_{\delta(-\vk)}
\nonumber
\\
&=&
V^2 \frac{P(q_i)P(|\vk-\vq_i|)}{|f(\vq_i, \vk-\vq_i;-\vk)|^2}\( \delta^{\rm K}_{\vq_j, \,-\vq_i} + \delta^{\rm K}_{\vq_j,\,-(\vk-\vq_i)} \)  
\nonumber
\\
&&+
V \frac{T\(\vq_i, \vk-\vq_i, \vq_j, -\vk-\vq_j\)}{f(\vq_i, \vk-\vq_i;-\vk) f(\vq_j, -\vk-\vq_j;\vk)} \,,
\label{eq:Cov_full}
\end{eqnarray}
where $T\(\vq_i, \vk-\vq_i, \vq_j, -\vk-\vq_j\)$ is the connected four-point function, or trispectrum, defined as follows:
\ba
T\(\vq_i, \vk-\vq_i, \vq_j, -\vk-\vq_j\) =& 
\<\delta(\vq_i)\delta(\vk-\vq_i)\delta(\vq_j)\delta(-\vk-\vq_j)\>
\vs
&- 
\<\delta(\vq_i)\delta(\vk-\vq_i)\>\<\delta(\vq_j)\delta(-\vk-\vq_j)\>
\vs
&-
\<\delta(\vq_i)\delta(\vq_j)\>\<\delta(\vk-\vq_i)\delta(-\vk-\vq_j)\>
\vs
&-
\<\delta(\vq_i)\delta(-\vk-\vq_j)\>\<\delta(\vk-\vq_i)\delta(\vq_j)\>\,.
\ea

After removing the duplicated contributions by imposing the inequality in \refeq{qrange}, the covariance matrix becomes
\bea
C_{ij}(\vq_i, \vq_j;\vk) =V^2 \frac{P(q_i)P(|\vk-\vq_i|) }{|f(\vq_i, \vk-\vq_i;-\vk)|^2}\delta^{\rm K}_{\vq_j, \,-\vq_i}
+
V \frac{T\(\vq_i, \vk-\vq_i, \vq_j, -\vk-\vq_j\)}{f(\vq_i, \vk-\vq_i;-\vk) f(\vq_j, -\vk-\vq_j;\vk)}\,.
\nonumber\\
\eea
From the covariance matrix, we compute the full noise power spectrum for the optimal quadratic estimator, \refeq{estimator}:
\bea
P_{\rm full}^N(k) 
=
P^N_G(k) + \bigg\{ \[P^N_G(k)\]^2 \sum_{ij} \frac{T\(\vq_i, \vk-\vq_i, \vq_j, -\vk-\vq_j\) f(\vq_i, \vk-\vq_i) f(\vq_j, -\vk-\vq_j)}{V^2 P(q_i) P(q_j) P(|\vk-\vq_i|) P(|-\vk-\vq_j|)} \bigg\}\,.
\nonumber\\
\label{eq:PNfull}
\eea

Note that the estimator \refeq{estimator} is optimal under the Gaussian assumption but may not remain optimal when the trispectrum contributes significantly to the covariance matrix. Since the relative contribution from the trispectrum rises toward small scales \citep{sefusatti2005galaxy,gualdi2022integrated}, we expect that the noise power spectrum $P^N_{\rm full}(k)$ would saturate beyond some $q_{\rm max}$.
Later, we verify the statements by comparing numerically the two noise power spectra $P^N_{\rm full}(k)$ and $P^N_G(k)$ for the two nonlinear density realizations: 2LPT \citep{Crocce2006Transients} and GridSPT \citep{Taruya2018Grid}.

\section{Leading-order clustering-fossil estimator and  nonlinear density fields}
\label{sec:implement}
The main goal of this paper is to assess the requirement for an accurate and unbiased clustering-fossil estimator in \refeq{estimator}. As for the simplest but realistic nonlinear density field, as shown in \refsec{spt2nd}, we study the clustering-fossil estimator using the second-order density perturbations as defined in the SPT (Standard Perturbation Theory) \citep{bernardeau2002large}. Specifically, we apply the quadratic clustering-fossil estimator to measure the long modes from the squeezed-limit bispectrum that encodes an off-diagonal power spectrum of the short modes.

Our approach is different from the previous studies \citep{Li2020reconstruction_a, Li2020new}, which tested the quadratic estimator based upon the second-order SPT (\refeq{kernel} below) against a few $N$-body simulation results, or tested the second-order EFT against dark matter halo fields 
\citep{darwish2021density}. In these studies, they have drawn affirmative conclusions about the possibility of reconstructing large-scale modes. At the same time, the quadratic estimator must fail beyond some $q_{\rm max}$, because the squeezed-limit bispectrum deviates from the tree-level SPT or EFT prediction. Also, the noise power spectrum of the reconstructed lone mode must be underestimated because the trispectrum contribution becomes important on small scales. Neglecting the trispectrum contribution would bias the power spectrum of reconstructed long modes. We cannot distinguish these two effects in $N$-body simulations. Also, we need multiple realizations to draw a robust statistical conclusion, but $N$-body simulations are too expensive for this purpose.
 
In contrast, our method based on the SPT allows a more systematic and in-depth study of the method because we can generate nonlinear density fields with a well-controlled nonlinear order. In this paper, we adopt a novel grid-based standard perturbation theory (GridSPT) method to generate the density fields up to second-order (\refsec{gridspt}). We then study the higher-order contribution's effect on the fossil estimator by comparing the estimated long mode from GridSPT realizations with the result from the second-order Lagrangian Perturbation Theory (2LPT) realizations (\refsec{2lpt}) and fourth-order GridSPT (\refsec{4thgridspt}). Both GridSPT and 2LPT are fast enough to generate each realization in less than one minute.

We have performed the comparison study at two redshifts ($z=1$ and $z=0$). For all studies, we use the simulation box of $V=(1000~{\rm Mpc}/h)^3$, and we adopt the following cosmological parameters \citep{ade2016planck}: $\Omega_{b0}h^2 = 0.022307$, $\Omega_{c0}h^2 = 0.11865$, $h = 0.6778$, $n_s = 0.9672$, $\Omega_{\nu0}h^2 = 0.000638$, $\Omega_{\Lambda0} = 0.69179$, ${\cal A}_s = 2.147\times 10^{-9}$ and $\sigma_8 = 0.8166$. We use CAMB \citep{Lewis1999CAMB} to generate the input linear power spectrum of GridSPT and 2LPT.

\subsection{The leading-order matter bispectrum from standard perturbation theory}
\label{sec:spt2nd}
Constructing the quadratic clustering-fossil estimator requires the squeeze-limit bispectrum to set the fossil kernel $f(\vk_1,\vk_2;\vk)$ in \refeq{estimator}. 
In SPT, the leading-order matter bispectrum from nonlinear gravity is given as 
\bea
B(k_1,k_2,k_3)
=
2 F_{2}(\vk_1,\vk_2) \Plin(k_1)\Plin(k_2) + (2~{\rm cyclic})\,,
\eea
with the second-order SPT kernel
\bea
F_{2}(\vk_1,\vk_2)
=
\frac57 + \frac27\(\frac{\vk_1\cdot\vk_2}{k_1k_2}\)^2 + \frac12\frac{\vk_1\cdot\vk_2}{k_1k_2}\(\frac{k_1}{k_2}+\frac{k_2}{k_1}\)
\eea
and the linear matter power spectrum $\Plin(k)$. Note that the kernel $F_2(\vk_1,\vk_2)$ only depends on the three wavenumbers $(k_1,k_2,k_3)$ because of the statistical homogeneity and isotropy.

The clustering-fossil estimator facilitates the squeezed limit of the bispectrum:
\bea
\label{eq:treeBk_f}
B(k,k_1,k_2)
\xrightarrow{k\to0}
\[2F_2(\vk_1, \vk)P_{L}(k_1) + 2F_2(\vk_2, \vk)P_{L}(k_2)\] P_{L}(k)\,,
\eea
and we can read off the following fossil-kernel expression from \refeq{squeezedBk} as:
\bea
\label{eq:kernel}
f(\vk_1, \vk_2; \vk) = 2F_2(\vk_1,\vk)P_{L}(k_1) + 2F_2(\vk_2,\vk)P_{L}(k_2).
\eea

Note that the third term $2F_2(\vk_1,\vk_2)P_L(k_1)P_L(k_2)$ vanishes in the squeezed limit because $F(\vq,-\vq)=0$. 
For a finite $\vk$, this term indeed contributes to the statistics of the recovered long mode. For example, when computing the cross power spectrum between the recovered long modes (using \refeq{Wieq}) and the original one, the term yields the correction to the squeezed-limit expression in \refeq{Prm_leading} as
\bea
\label{eq:Prm}
P_{rm}(k) 
&=& \sum_i W_i \frac{B(\vq_i,\vk-\vq_i,-\vk)}{f(\vq_i,\vk-\vq_i;-\vk)}
\vs
&=& P_{L}(k) + 2P^N(k) \sum_i \frac{F_2(\vq_i, \vk - \vq_i) P_L(q_i) P_L(|\vk - \vq_i|)}{f(\vq_i, \vk-\vq_i;-\vk)}\,.
\eea

\subsection{Nonlinear density fields} 
\subsubsection{Second-order GridSPT}
\label{sec:gridspt}
The Grid-based standard perturbation theory (GridSPT) is a method of computing the $n$-th order SPT density and velocity fields from the recursion relation. Standard Perturbation Theory \citep{Vishniac:1983,Fry:1984,Goroff/etal:1986,Suto/Sasaki:1991,Makino/etal:1992,Jain/Bertschinger:1994,Scoccimarro/Frieman:1996,jeong2006perturbation} approximates the evolution as a pressure-less fluid with the following evolution equations for the density contrast $\delta({\bm x},\tau)\equiv\rho_m({\bm x},\tau)/\bar{\rho}_m(\tau)-1$ and the peculiar velocity ${\bm v}({\bm x},\tau)$:
\begin{eqnarray}
&&\dot{\delta} + \nabla\cdot\left[(1+\delta){\bm v}\right] = 0 \,, \label{eq:continuity}
\\
&&\dot{\bm v} + ({\bm v}\cdot\nabla){\bm v} + \frac{\dot a}{a}{\bm v} = - \nabla\Phi\,, \label{eq:euler}
\end{eqnarray}
along with the Poisson equation:
\be 
\nabla^2 \Phi = 4\pi G \bar{\rho}_\mathrm{m}\, a^2 \delta\,. \label{eq:poisson}
\ee
Here, dot represents the conformal-time derivative, $d\tau=dt/a$ with $a(t)$ being the scale factor and $t$ being the cosmic time, $\nabla$ is comoving-coordinate derivative, $\bar{\rho}_\mathrm{m}$ is the mean matter density, and $\Phi$ is the peculiar gravitational potential. Note that we omit the spacetime coordinate in equations to avoid clutter. 

The set of equations describes the non-relativistic-matter (cold-dark matter and baryon) fluid on scales larger than the baryonic Jeans scale. Note that the nonlinearities in \refeqs{continuity}{euler} comes from the second-order terms such as $\nabla\cdot(\delta{\bm v})$ and $({\bm v}\cdot\nabla){\bm v}$. In SPT, we solve \refeqs{continuity}{poisson} by expanding the nonlinear density contrast $\delta$ and velocity-gradient field $\theta\equiv-\nabla\cdot{\bm v}/(aHf)$ as
\be
\delta({\bm x},\tau)=\sum_n [D(\tau)]^n\delta^{(n)}({\bm x})\,,~~
\theta({\bm x},\tau)=\sum_n [D(\tau)]^n\theta^{(n)}({\bm x})\,,
\label{eq:def_delta}
\ee
where $D(\tau)$ is the linear growth factor.

For a given realization of the linear density field on regular grid points, the GridSPT \citep{Taruya2018Grid} provides a way to compute the matter density field $\delta$ and the velocity field ${\bm v}$ of LSS perturbatively by solving the fluid equations [\refeqs{continuity}{poisson}], which becomes the recursion relation as:
\begin{eqnarray}
\left(
\begin{array}{c}
    \delta^{(n)}(\vx)\\
    \theta^{(n)}(\vx)
\end{array}
\right)
&=&
\frac{1}{(2n+3)(n-1)}
\left(
\begin{array}{ll}
    2n+1 ~~& 1\\
    3    ~~& n
\end{array}
\right) 
\sum_{m=1}^{n-1}
\left(
\begin{array}{c}
    \nabla\cdot\left(\delta^{(m)}{\bm v}^{(n-m)}\right)\\
    \nabla^2\left({\bm v}^{(n-m)}\cdot{\bm v}^{(m)}\right)
\end{array}
\right)\,.
\label{eq:GridSPTrecursion}
\end{eqnarray}
From a given set of linear density field, $\delta_1=\theta_1=\delta_L$, we can use \refeq{GridSPTrecursion} to compute the nonlinear density $\delta^{(n)}$ and velocity-gradient $\theta^{(n)}$ fields order by order manner. Using the FFT (Fast Fourier Transform) to evaluate the $\nabla$ operators on the right-hand side, the GridSPT enables us to quickly generate the $n$th order quantities. 

We use the second-order GridSPT, which contains first- and second-order density contrast, to test the fossil estimator. The squeezed bispectrum in the second-order GridSPT strictly follows the tree-level bispectrum in SPT so the constructed estimator can fully capture the coupling between long and short modes. Therefore, the second-order GridSPT is an ideal tool to conduct a proof-of-concept study of the fossil estimator without any uncontrolled systematic error.

We compute the GridSPT density fields on the $N_{\rm grid} = 512^3$ grids. We adopt the empirical cutoff $k_1 = 1.0\, h$/Mpc for the first-order density contrast and $k_2 = 1.33\, h$/Mpc for higher-order density contrasts used in \citep{Taruya2018Grid}.

\subsubsection{Second-order Lagrangian perturbation theory (2LPT)}
\label{sec:2lpt}
Lagrangian perturbation theory (LPT) is an alternative to the SPT in modeling the nonlinear density fields. The fundamental object in LPT is the displacement field $\bPsi(\vq,\tau)$ from the regular Lagrangian position $\vq$, which makes the Eulerian position $\vx$ as
\be
\vx(\vq,\tau) = \vq + \bPsi(\vq,\tau)\,.
\label{eq:LPTdisp}
\ee
We can obtain the LPT solutions by solving the equation of motion in an expanding universe
\be
\ddot{\vx} + \frac{\dot{a}}{a}\dot{\vx} = - \nabla_{\vx}\Phi\,
\label{eq:LPTeom}
\ee
for an irrotational displacement field $\nabla\times\bPsi=0$ perturbatively.
The second-order solution for the displacement field is given as (see, for example, appendix E of \cite{Jeong:thesis} for a full derivation)
\be
\bPsi(\vq,\tau)
=
-\nabla_{\vq}\phi^{(1)}(\vq,\tau)
+\nabla_{\vq}\phi^{(2)}(\vq,\tau)\,,
\label{eq:Psi2nd}
\ee
where the linear Lagrangian potential is related to the linear density contrast as
\be
\nabla_{\vq}^2\phi^{(1)}(\vq,\tau) = \delta^{(1)}(\vx,\tau)\,,
\ee
and the second-order Lagrangian potential is related to $\phi^{(1)}$ as
\be
\nabla_{\vq}^2\phi^{(2)}(\vq,\tau)
=
-\frac{D_2(\tau)}{D(\tau)^2}
\sum_{i>j}
\left\{
\phi^{(1)}_{,ii}(\vq,\tau)
\phi^{(1)}_{,jj}(\vq,\tau)
-
\left[
\phi^{(1)}_{,ij}(\vq,\tau)
\right]^2
\right\}\,.
\ee
Here, $D_2(\tau)$ is the solution of the following differential equation:
\begin{equation}
\ddot{D}_2(\tau) + \frac{\dot{a}}{a}\dot{D}_2(\tau) - \frac{3}{2}\(\frac{\dot{a}}{a}\)^2\[\Omega_m(\tau)D_2(\tau) - D^2(\tau)\] = 0\,,
\end{equation}
and $D_2 = -3/7D^2(\tau)$ for the Einstein de-Sitter (spatially flat and matter-dominated) universe.

The 2LPT (seond-order LPT) prescription is to displace regularly spaced particles using \refeq{LPTdisp} and \refeq{Psi2nd}. Since the nonlinearities are modeled by particle displacement, the resulting density contrast, while agreeing with the SPT prediction to second order, contains a myriad of higher-order nonlinear contributions (see, e.g., Refs.~\cite{Crocce2006Transients} and \cite{McCullagh2016Toward}). The bispectrum of the 2LPT density field, therefore, deviates from the tree-level predictions in \refeq{treeBk_f}, particularly on small scales. By applying the same fossil estimator using the kernel given in \refeq{kernel} to the 2LPT density fields, we can test the behavior of the fossil estimator when ignoring the higher-order nonlinear couplings in data.

We implement the 2LPT by using the displacement of $512^3$ Lagrangian particles in the cubic box of volume (1000 Mpc/$h$)$^3$. We then estimate the density with $N_{\rm grid} = 256^3$ grids and preserve the density modes of wavenumber $k < k_{\rm Nyq}/2$ to avoid the aliasing effect \citep{Jeong:thesis}.

\subsubsection{Fourth-order GridSPT}
\label{sec:4thgridspt}

While we test the effect of neglecting higher-order nonlinear couplings by comparing the second-order GridSPT and 2LPT, the density fields in these two toy cases are far from reality on small scales. For example, the nonlinearity in the density field is too strong in second-order GridSPT \citep{wang2022nEPT} while too weak in 2LPT \citep{Taruya2018Grid} compared to $N$-body simulations. To overcome this, we also test the performance of the tree-level fossil estimator in a more realistic density field using the fourth-order GridSPT. The power spectrum and bispectrum in the fourth-order GridSPT can be accurately modeled by the complete one-loop power spectrum and one-loop bispectrum in SPT, which are closer to the $N$-body result than either second-order GridSPT or 2LPT \citep{wang2022nEPT}. As for the $q_{\rm max}(z)$, the smallest scale to be included in the reconstruction at each redshift, we use the result of Ref.~\cite{tomlinson2022Bk}, which has measured the maximum wavenumber below which the tree-level bispectrum accurately model the nonlinear bispectrum from a suite of $N$-body simulations.

\section{Results}
\label{sec:results}
We present the results of the analysis in reconstructing the long modes by applying the tree-level fossil estimator to the following three nonlinear density fields: second-order and fourth-order GridSPT and 2LPT.

\subsection{The tree-level fossil estimator on the 2LPT density field}
\label{sec:2lptresult}
First, we applied the tree-level fossil estimator to the 2LPT density fields to reconstruct the long modes.

\subsubsection{The reconstructed long modes vs. the original modes}
\begin{figure}
    \centering
    \includegraphics[width = 0.9\columnwidth]{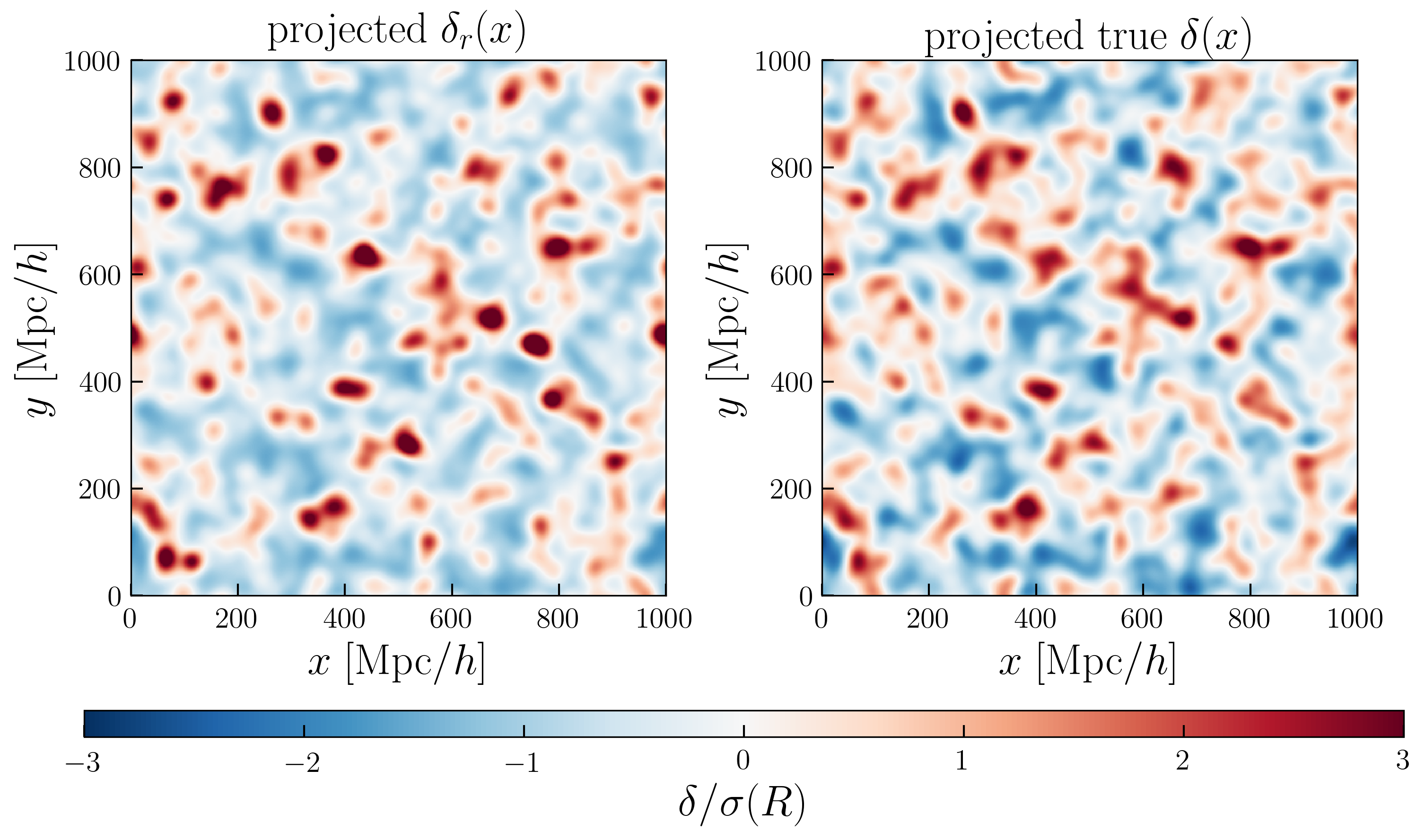}
    \caption{The two-dimensional (the 2 $h/$Mpc slice on the $x$-$y$ ($z=0$) plane) morphology of the reconstructed density field ({\it Left}) and the 2LPT density fields ({\it Right}) at $z = 1$. We smooth all fields using the spherical Gaussian filter with the radius $R = 15\;h$/Mpc. The maximum wavenumber of modes used for reconstruction is $q_{\max} = 0.4 \,h/$Mpc. As indicated by the color bar, both are normalized to their own variance.}
    \label{fig:Morphologyz1}
\end{figure}
\begin{figure}
    \centering
     {\includegraphics[width = 0.5\textwidth]{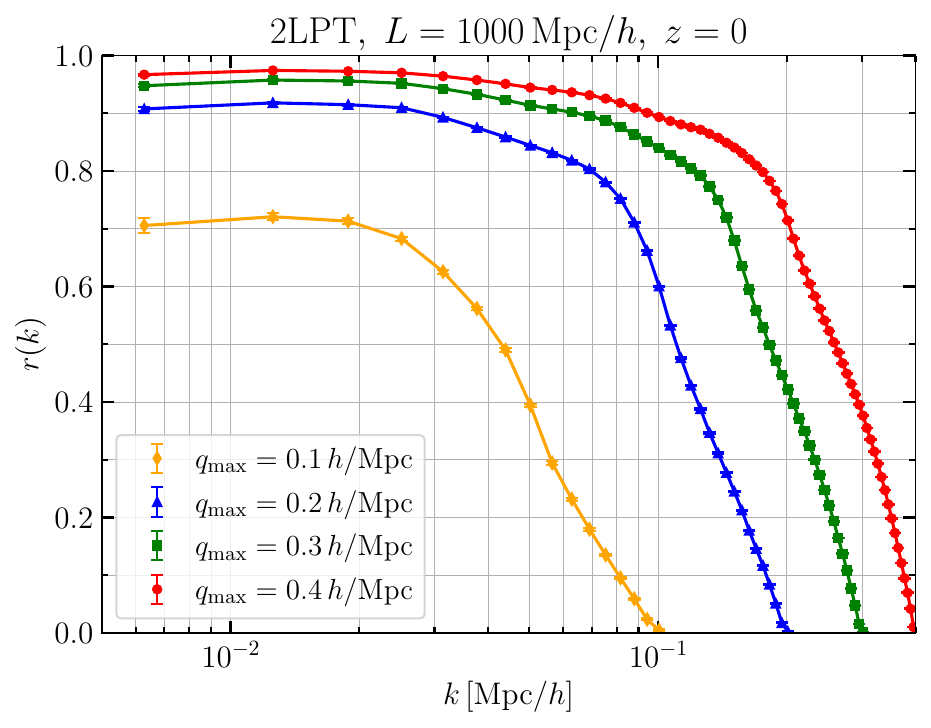}}%
     {\includegraphics[width = 0.5\textwidth]{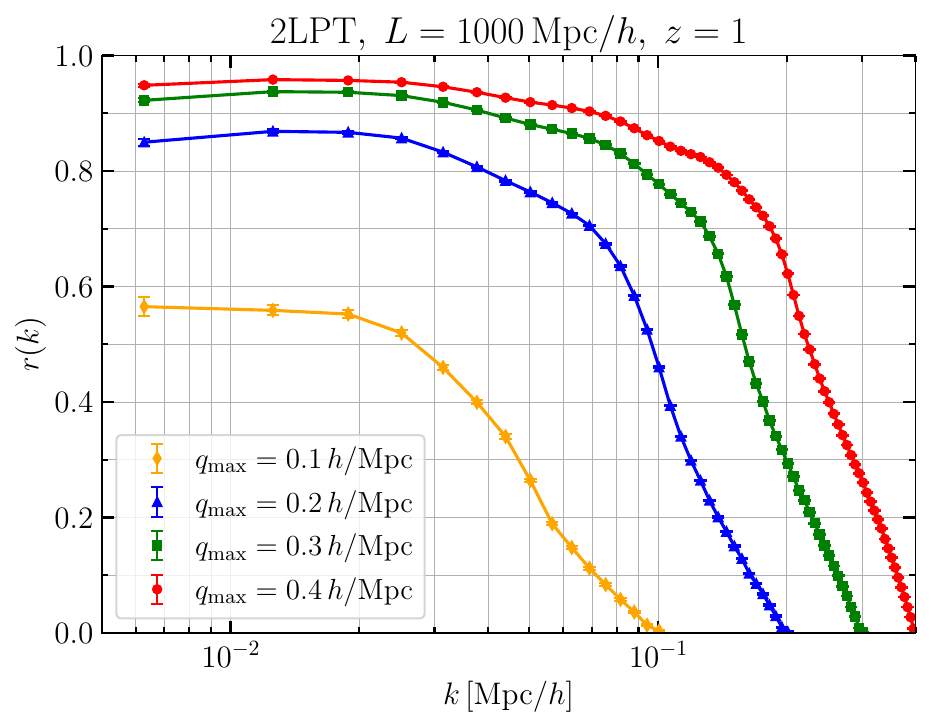}}%
    \caption{The cross-correlation coefficient, $r(k)$ defined in \refeq{rk}, between the reconstructed and true long modes in 2LPT simulations at $z=0$ (left) and $z=1$ (right). Each line shows the result from different maximum wavenumber of short modes for reconstruction, $q_\max$ = $0.1$, $0.2$, $0.3$, and $0.4~h/$Mpc. The error bars indicate the standard deviation of the mean measured from 100 2LPT realizations.}
    \label{fig:Rkz0z1}
\end{figure}

In \reffig{Morphologyz1}, we compare the reconstructed (left panel) and true (right panel) large-scale density field. We show the projected density distribution at $z=1$ in the $x$-$y$ plane of thickness 2 Mpc$/h$. We use short modes up to $q_\max=$0.4 $h/$Mpc for reconstruction, and we smooth both fields with a Gaussian filter of radius $R$ = 15 Mpc$/h$. In general, the reconstructed density field resembles the morphology of the original density field, but we can clearly see the visible differences between the two. This is most apparent for small-scale features in that some features in the original 2LPT field are missing in the reconstructed field, which implies that the recovered mode is more noise-dominated on smaller scales. 

For a more quantitative comparison, we compute the cross-correlation coefficients between the recovered and original long modes, and the result is shown in \reffig{Rkz0z1} for the results at redshifts $z = 0$ (left) and $z =1$ (right). At each redshift, we show four different results corresponding to four different values of $q_{\rm max}=0.1$, $0.2$, $0.3$, and $0.4~h/$ Mpc. 

First of all, the cross-correlation coefficient quickly drops as the long-mode wavenumber $k$ approaches $q_\max$. That is because there are fewer and fewer short modes for reconstruction as $k$ approaches $q_{\rm max}$. Then the recovered long modes are dominated by noise and correlate weakly with the original long modes. This is consistent with the lack of small-scale features we observe in the morphological comparison in \reffig{Morphologyz1}. 

We also notice that the cross-correlation coefficients increase with $q_{\rm max}$, with an especially large improvement from $q_\max = 0.1\; h/$Mpc to 0.2 $h/$Mpc. With $q_\max = 0.4\;h$/Mpc, the cross-correlation coefficient reaches 0.95 on the large-scale end. This implies that the phase of the recovered long mode from the fossil estimator is highly correlated with the true long mode. Clearly, including more short modes leads to a more accurate reconstruction of the phase of the long modes. 

Comparing the two panels, for a fixed $q_{\rm max}$, the cross-correlation coefficient at $z=0$ is larger than $z=1$ for all four $q_{\rm max}$ cases. As derived in \refeq{Rk}, the cross-correlation coefficient is close to unity when $P_G^N/P_L\ll1$. Since the noise power spectrum $P_G^N$ on large scales only weakly depends on the redshift [\refeq{PN_G} and \refeq{kernel}] while the amplitude of linear power spectrum grows in time, the cross-correlation coefficient is closer to unity at lower redshifts. In other words, a higher signal-to-noise ratio of the power spectrum leads to a larger cross-correlation coefficient at lower redshifts. This phenomenon can also be interpreted as a consequence of the stronger nonlinear coupling between the long and short modes at lower redshift. That is, at higher redshifts, we can reach the same level of the cross-correlation coefficient only with an increasing dynamic range of the short modes used for the reconstruction.

\subsubsection{Signal-to-noise ratio}
\label{sec:noisepk}
\begin{figure}
    \centering
    {\includegraphics[width = 0.49\textwidth]{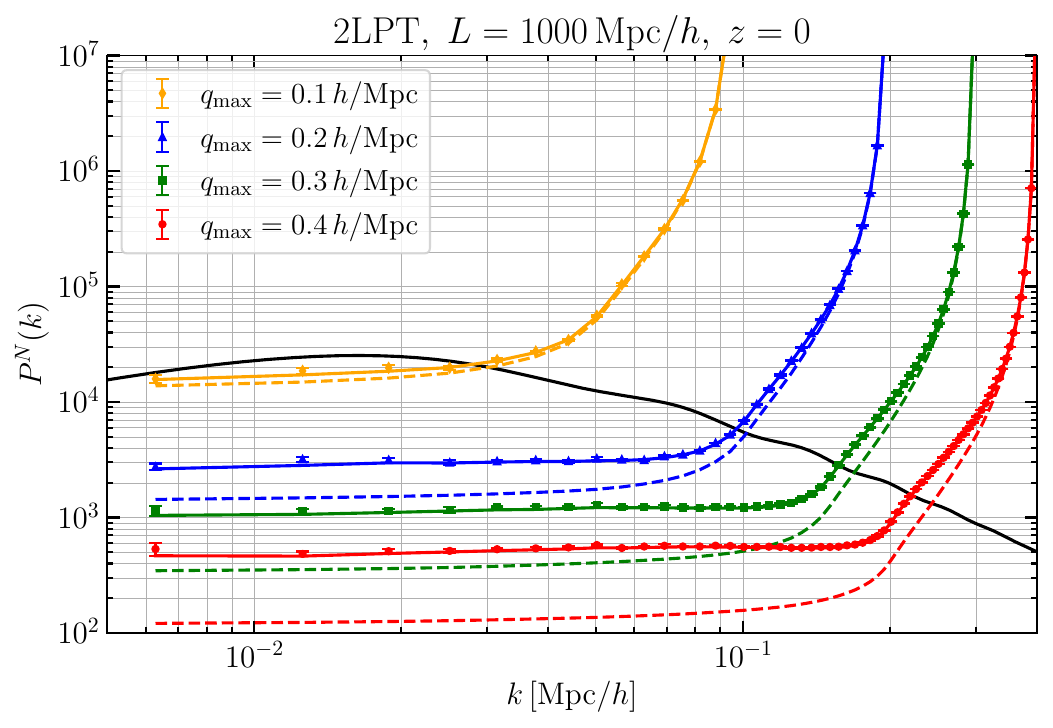}}
    {\includegraphics[width = 0.49\textwidth]{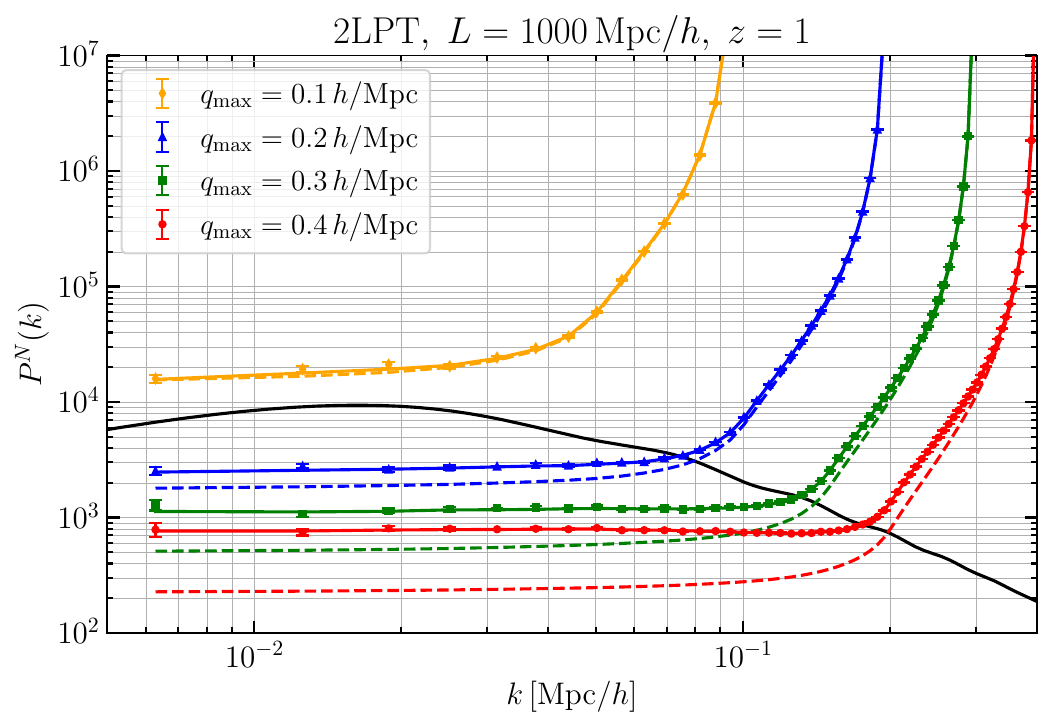}}
    \caption{The significance of the long-mode reconstruction at $z=0$ (left panel) and $z=1$(right panel). For both panels, we show the linear matter power spectrum (signal) as a solid black line and the noise power spectrum with four different $q_\max=0.1$, $0.2$, $0.3$, and $0.4~h/$Mpc as different colors: measured from ten 2LPT realizations (dots with error bars), Gaussian estimate [\refeq{PN_G}] (dashed lines), and non-Gaussian estimate, full expressions in \refeq{PNfull} (solid lines).}
    \label{fig:PNz0z1}
\end{figure}
To test the statistical significance of the reconstruction, we measure the noise power spectrum of the reconstructed long modes and show them in \reffig{PNz0z1} for two redshifts $z = 0$ (left) and $z = 1$ (right), and for four different maximum short mode wavenumber, $q_{\rm max}$. For each case, we compare the measured noise power spectrum (data points with errorbar) with the theoretical estimation from Gaussian assumption \refeq{PN_G} (dashed line) and the full computation [\refeq{PNfull}] including the trispectrum contribution (solid line). We find that while the full noise power spectrum estimate captures the measurement quite well, the Gaussian approximation always underestimates the noise. As coming from the non-Gaussian trispectrum, the discrepancy is most apparent from cases including more small-scale contributions or increasing $q_{\rm max}$. 

The signal-to-noise ratio of the long-mode reconstruction significantly improves as increasing $q_{\rm max}$, because more short-modes are used for the reconstruction. At the same time, including short modes in the nonlinear scales deviates the noise power spectrum from the Gaussian approximated one. Such a deviation does not merely cause an underestimation of the errorbar, because, as shown in \refeq{Prrk}, estimating the long-mode power spectrum requires subtracting the noise power spectrum. That is, underestimation of the noise power spectrum leads to the systematic overestimation bias of the long-mode power spectrum. \refsec{recoveredPk} demonstrates this point explicitly.

\subsubsection{The power spectrum of the recovered long modes}
\label{sec:recoveredPk}
\begin{figure}
    \centering
    \includegraphics[width = 0.49 \columnwidth]{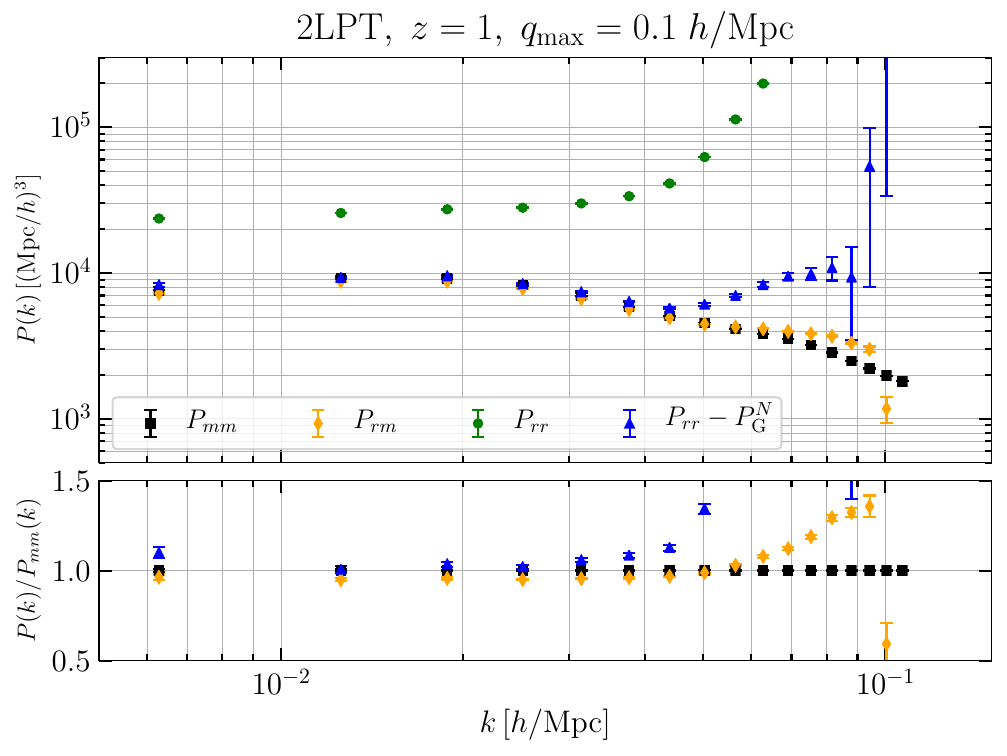}
    \includegraphics[width = 0.49 \columnwidth]{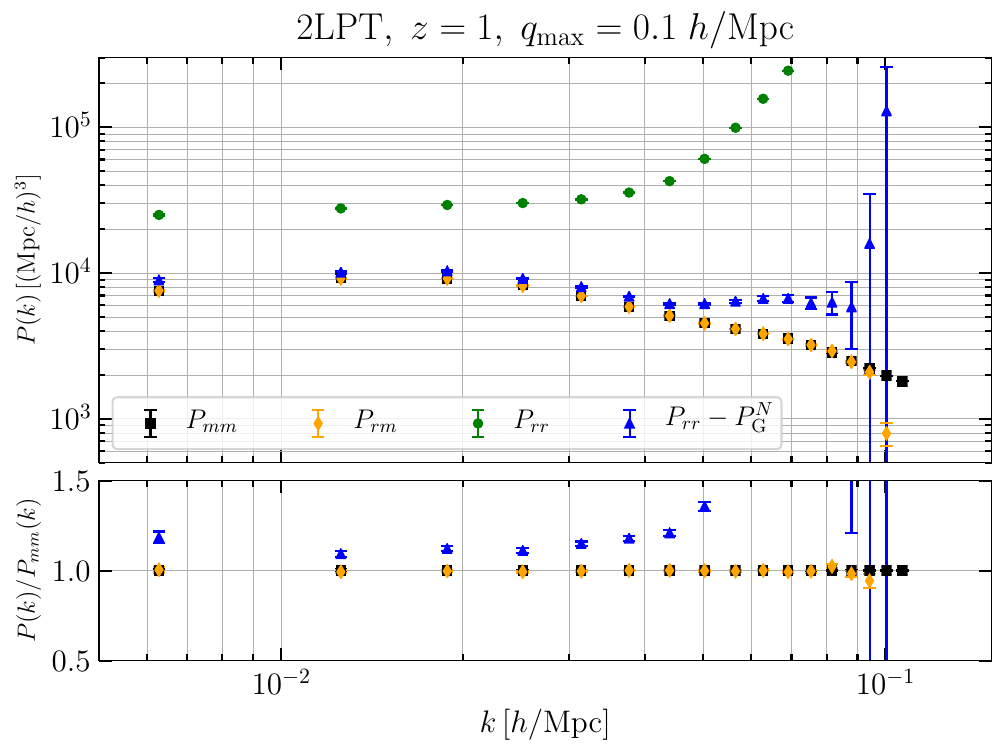}
    \caption{The ensemble mean of the power spectrum from 1,000 2LPT simulations at $z = 1$, using $q_{\max} = 0.1\,h$/Mpc. We highlight the importance of the fossil kernel by comparing the fossil kernel $f(\vk_1,\vk_2;\vk)$ from the tree-level bispectrum ({\it Left}) and from the bispectrum and power spectrum measured from an average of 10,000 2LPT realizations ({\it right}). For both panels, black (sqaures), orange (diamonds), blue (triangles), and green (dots) are, respectively, the true matter power spectrum, the cross power spectrum between the recovered and the original field, and the recovered power spectrum subtracting the noise calculated from Gaussian approximation. We also show the ratio of each power spectrum to the $P_{mm}(k)$ in the bottom panels.
    \label{fig:2LPTPk}
    }
\end{figure}

In \reffig{2LPTPk}, we show the ensemble mean of the power spectra from the long modes, both auto and cross-correlation with the original mode, reconstructed from the 1,000 2LPT realizations at $z = 1$ (with $q_{\rm max} = 0.1\, h$/Mpc) and compare them with the power spectrum $P_{mm}(k)$ of the original long mode. A perfect reconstruction would yield a perfect match.

The left panel of \reffig{2LPTPk} shows the reconstruction with the tree-level fossil kernel $f(\vk_1,\vk_2;\vk)$ in \refeq{kernel} coming from the tree-level bispectrum, and the right panel shows the results with the measured fossil kernel. We measured the fossil kernel by taking the ratio between the average of the bispectrum and the average long-mode power spectrum using the 10,000 2LPT simulations.

The cross-power spectrum $P_{rm}(k)$ from the tree-level fossil kernel (left panel) deviates from $P_{mm}(k)$, while that from the measured fossil kernel (right panel) is on top of the $P_{mm}(k)$. It is because the 2LPT bispectrum, even at $q_{\rm max}=0.1~h/{\rm Mpc}$ at $z=1$, deviates from the tree-level prediction. This result highlights the importance of having an accurate fossil kernel for the reconstruction. \refeq{Prm} shows that the cross power spectrum $P_{rm}$ is determined by the ratio between the true bispectrum and the fossil kernel, which is equal to the linear matter power spectrum $P_{L}$ on large scales. 
The high-$k$ deviation of $P_{rm}$ from $P_{mm}$ is due to the correlation between the second-order long mode and two first-order short modes (the second term in \refeq{Prm}).

The auto power spectrum of the recovered modes $P_{rr}$ exceeds $P_{mm}$ due to the presence of the noise power spectrum (see \refeq{Prrk}). Therefore, $P_{rr}$ increase rapidly when $k$ approaches $q_\max$. This is also consistent with the quick drop of the cross-correlation coefficient at high $k$ as we showed in \reffig{Rkz0z1}. Subtracting the noise power spectrum $P_G^N$ estimated using the Gaussian approximation reduces them closer to $P_{mm}(k)$, but still $P_{rr}-P_G^N$ disagrees with $P_{mm}$. As we shall show in \refsec{2gridSPT}, this is due to the trispectrum contribution to the noise power spectrum.

\subsection{The ideal toy: reconstruction from the second-order GridSPT}
\label{sec:2gridSPT}
\begin{figure}
    \centering
    \includegraphics[width = 0.5 \columnwidth]{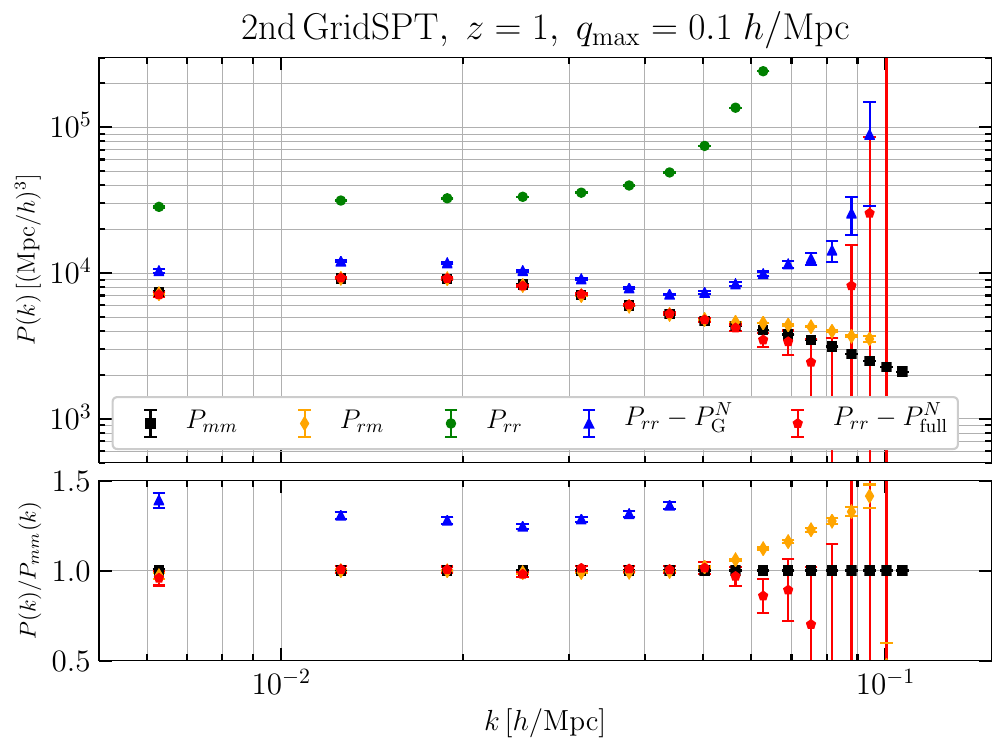}
    \caption{The reconstructed long mode power spectrum when applying the fossil estimator to the second-order GridSPT density field. The symbols are the same as \reffig{2LPTPk} except for the red pentagon, which shows the recovered power spectrum with the full noise subtracted. The recovered long-mode power spectrum matches the ground truth (black square) only after subtracting the noise power spectrum taking into account the non-Gaussian covariance.
    \label{fig:Pk}}
\end{figure}

In \refsec{2lpt}, we show that using an accurate squeezed-limit bispectrum for the fossil kernel is essential to get the correct cross-correlation power spectrum. At the same time, the estimated auto-correlation of the recovered long mode deviates from that of the original field when subtracting the noise power spectrum estimated using the Gaussian approximation. To show the proof-of-concept case where we can recover both cross-correlation and auto power spectrum of the long mode, we apply the fossil estimator to the second-order GridSPT realizations. The second-order GridSPT realizations contain the nonlinear density field up to the second order so that we can estimate all relevant correlations.

\reffig{Pk} show the result of the reconstruction analysis using 1,000 second-order GridSPT realizations with $q_{\max}=0.1\,h$/Mpc. Just like in \refsec{2lpt}, when estimating the noise power spectrum with Gaussian approximation, the recovered power spectrum (the blue triangles) is about 30 percent higher than $P_{mm}$. We reach the agreement (red pentagon symbols) only after including the full covariance matrix with the trispectrum contribution. To do so, we measure the off-diagonal covariance matrix numerically from 10,000 realizations of second-order GridSPT and calculate the full noise power spectrum $P^N_{\rm full}$ according to \refeq{PNfull}. 

Therefore, it is essential to model the correct noise power spectrum, including the off-diagonal terms in the covariance matrix or trispectrum. Neglecting these terms not only underestimates the uncertainties in the estimated long-mode power spectrum, but also introduces systematic bias in the power spectrum of recovered long modes.

\subsection{A more realistic toy: reconstruction with the fourth-order GridSPT}
Thus far, we have reconstructed the long modes from the controlled toy nonlinear density fields generated from the 2LPT (\refsec{2lpt}) and the second-order GridSPT (\refsec{2gridSPT}), from which we find that we have to use an accurate fossil kernel and to incorporate the non-Gaussian covariance (including the trispectrum) in order to achieve an unbiased reconstruction of the long modes. Although useful for the analysis, of course, none of the test nonlinear density fields is close to the real cosmic density field. In this section, we study the limitations of the tree-level fossil estimator by using the fourth-order GridSPT density field.

\subsubsection{The reach of the tree-level fossil estimator}
\label{sec:limittree}
\begin{figure}
    \centering
    \includegraphics[width = 0.5 \columnwidth]{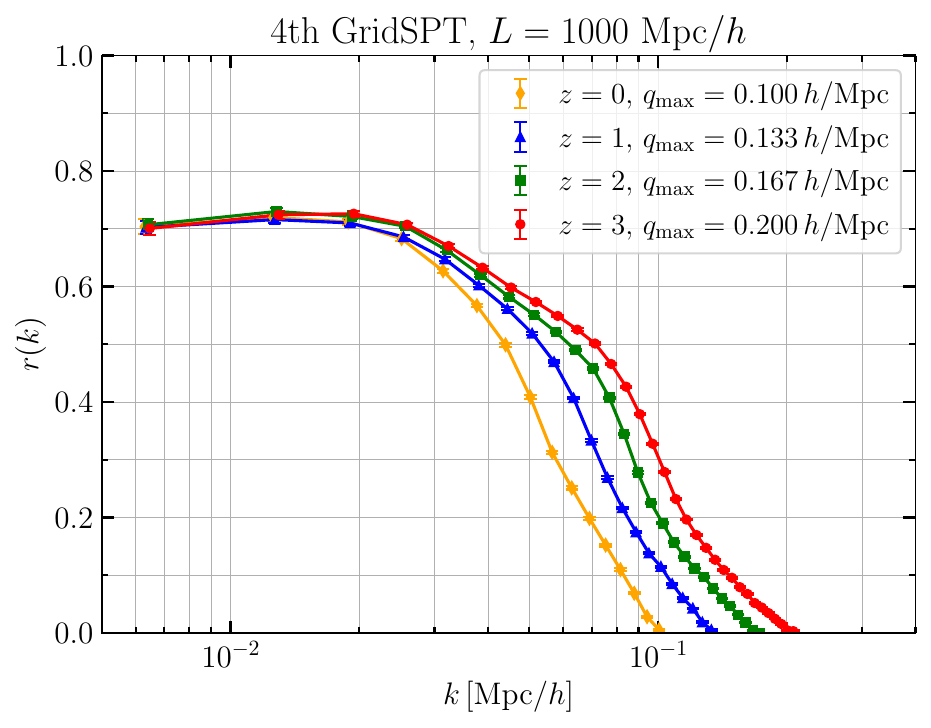}
    \caption{The cross-correlation coefficient between the recovered field and the original field constructed using four-order GridSPT realizations. For the reconstruction, we use four different $q_{\max} = 0.1,\,0.133,\,0.167,\,0.2\,h$/Mpc, respectively, at $z=0$, $1$, $2$, $3$ such that the squeezed bispectrum works within 2\% accuracy, as determined by \cite{tomlinson2022Bk}. The error bars indicate the standard deviation of the mean measured from 100 GridSPT realizations.}
    \label{fig:rGridn4_z0123}
\end{figure}
\begin{figure}
    \centering
    {\includegraphics[width = 0.48 \textwidth]{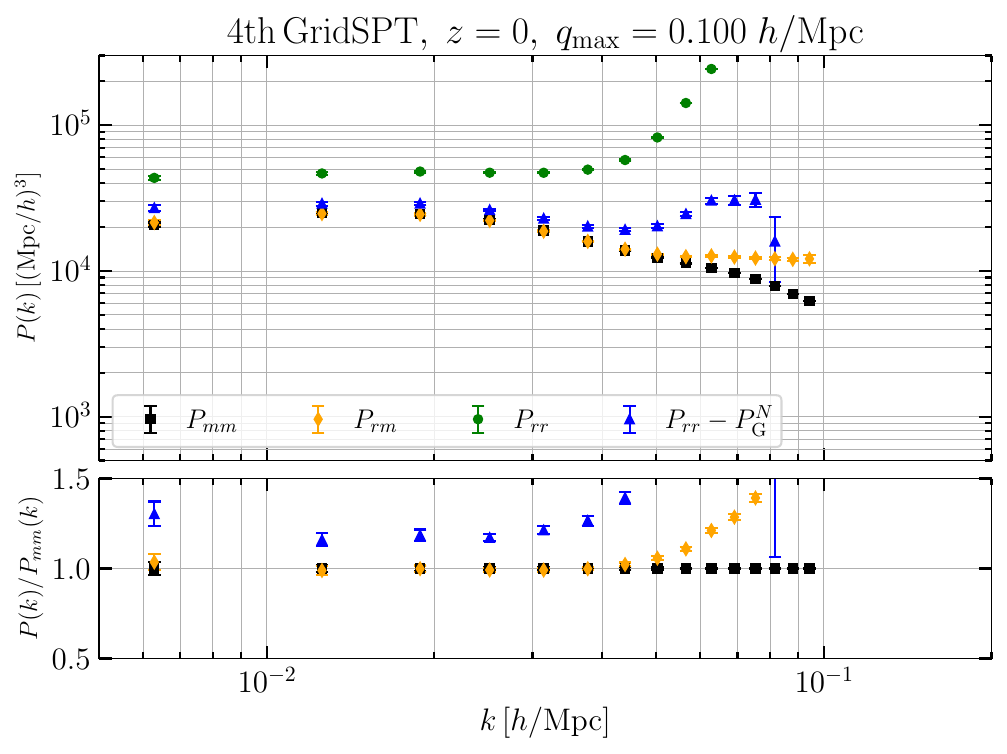}}
    {\includegraphics[width = 0.48 \textwidth]{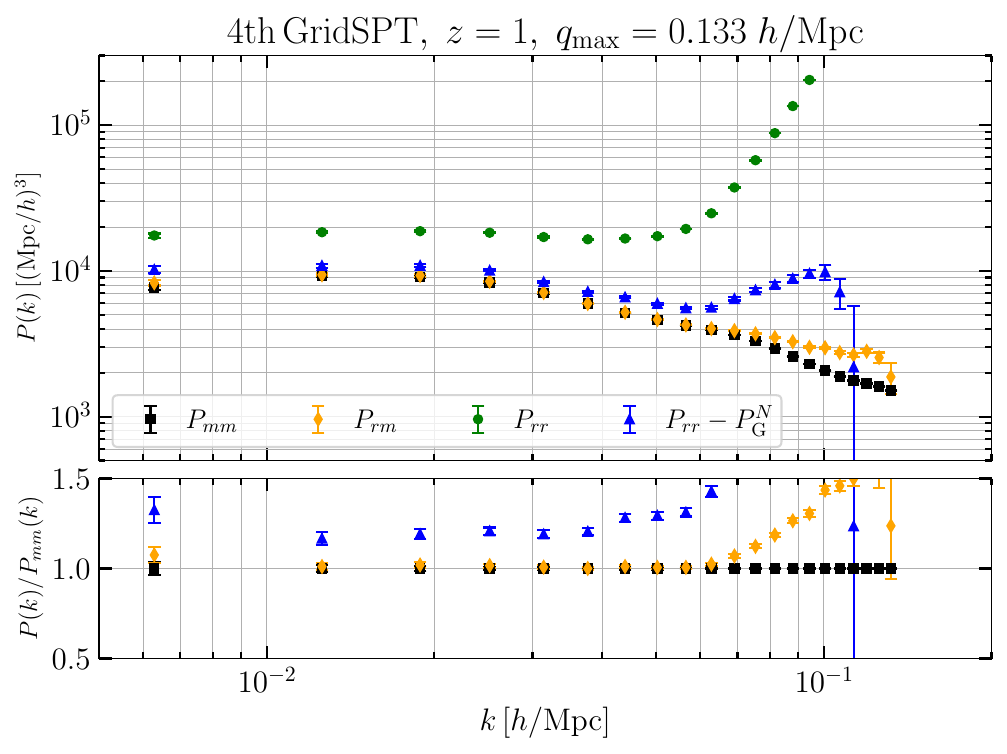}}
    {\includegraphics[width = 0.48 \textwidth]{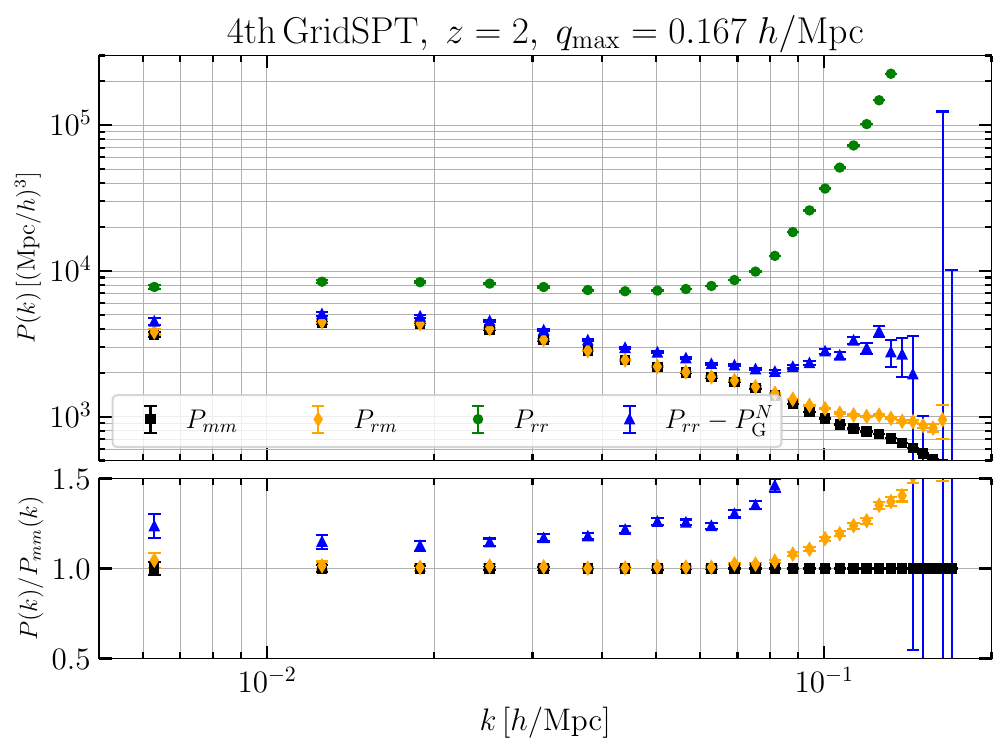}}
    {\includegraphics[width = 0.48 \textwidth]{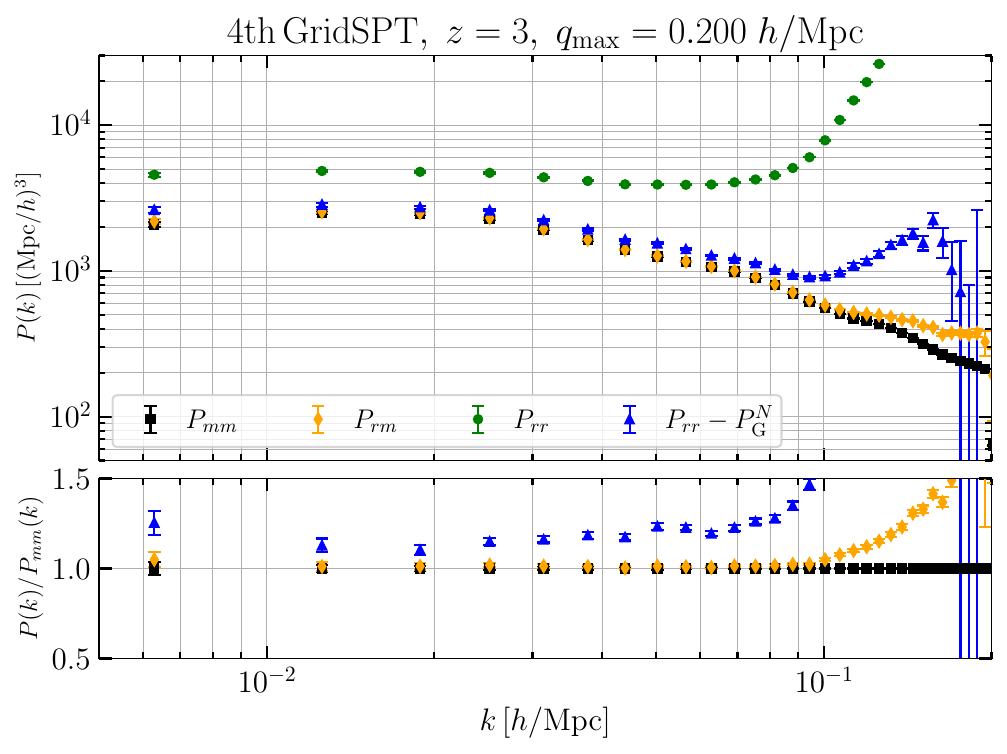}}
    \caption{The ensemble mean of the power spectrum measured from 1,000 GridSPT realizations at $z = 0$, $1$, $2$, and $3$, using the $q_{\max}$ determined as one-loop bispectrum models the squeezed-limit with 2~\% accuracy \cite{tomlinson2022Bk}. The blue triangles are the recovered power spectrum subtracting the noise calculated from the Gaussian approximation. The orange diamonds are the cross-power spectrum between the recovered and the original field. Symbols are identical to \reffig{2LPTPk}. For all cases, the bottom panel shows the ratio between the power spectra and the true matter power spectrum $P_{mm}$.}
    \label{fig:Gridn4_z0}
\end{figure}

We apply the tree-level fossil estimator in fourth-order GridSPT, using $q_{\rm max} = 0.1$, $0.133$, $0.167$, $0.2~h$/Mpc, respectively, at $z = 0,1,2,3$, which are the highest wavenumber where the tree-level bispectrum models the measured squeezed bispectrum in $N$-body simulations within the $2\%$ accuracy \cite{tomlinson2022Bk}. That is, the reconstruction must be unbiased within these dynamic ranges.

In \reffig{rGridn4_z0123}, we present the best cross-correlation coefficient that the estimator can achieve at $z = 0,\,1,\,2,\,3$. For all four redshifts on large scales, the cross-correlation coefficients reach $r = 0.7$, which indicates the best signal-to-noise ratios of the recovered long mode based on tree-level bispectrum theory. 

\reffig{Gridn4_z0} presents the measured auto- and cross- power spectra at the four redshifts. One notable feature is that the cross-power spectra match the true matter power spectra on large scales. This fact ensures that the tree-level fossil estimator is unbiased in the dynamical ranges. To enhance the signal-to-noise ratio of the reconstructed mode, we need to find a more accurate bispectrum model, e.g. one-loop SPT bispectrum, such that the estimator can safely include more short modes on smaller scales without biasing the recovered long mode.

\subsubsection{The limit of the tree-level fossil estimator}
\label{sec:limit}
\begin{figure}
    \centering
    \includegraphics[width = 0.5\columnwidth]{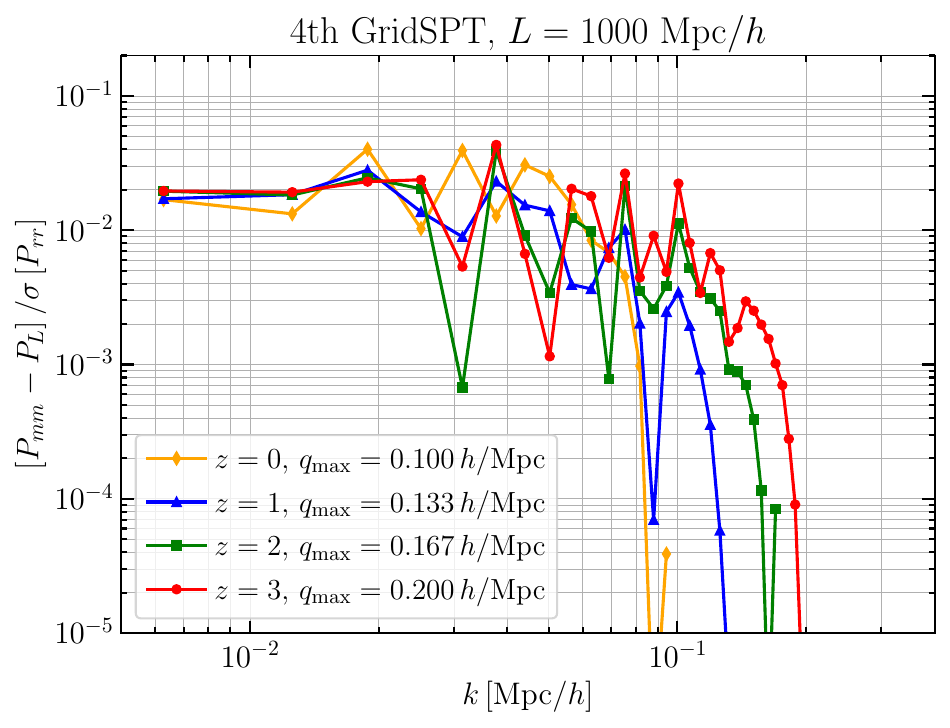}
    \caption{The ratio between the neglected nonlinear correction to the reconstructed matter power spectrum and the cosmic variance of the recovered auto power spectrum at $z=0, 1, 2, 3$. The results are measured in 100 fourth-order GridSPT realizations. We use the same $q_{\rm max}$ values as in \reffig{rGridn4_z0123}.}
    \label{fig:fossillimit}
\end{figure}

In \refsec{fossil}, we have proved that the fossil estimator is equivalent to the squeezed-limit bispectrum. Therefore, the fossil estimator reconstructs the linear-order density mode while the nonlinear part of the density mode is neglected. We test the scheme in this section by comparing the reconstructed power spectrum with the original density power spectrum which contains the nonlinear terms.

\reffig{fossillimit} shows the difference between the reconstructed power spectrum and the original long-mode power spectrum in a unit of the cosmic variance uncertainty $\sigma(P_{rr})$. We use 100 fourth-order GridSPT realizations for this analysis. For all ($z = 0, 1,2, 3$) redshifts. This plot shows that the nonlinear corrections are only a few percent of the cosmic variance at all four redshifts, which indicates the nonlinear correction of the recovered mode can be safely neglected with the range of $q$ and the simulation box size we use in this study. Note that the nonlinear correction would become important if we work with either a bigger simulation box or a larger dynamical range of $q$, because the former raises the number of modes while the latter suppresses the noise power spectrum in $P_{rr}$. In either case, the nonlinear correction could be comparable to the cosmic variance of $P_{rr}$. And we have to carefully pick out the range of wavenumbers that the fossil estimator works.

\section{Conclusion \& Discussion}
\label{sec:conclusion}

Here, we investigate the reconstruction of long-wavelength density modes based on the off-diagonal correlation between short-wavelength modes in the Fourier space, a phenomenon called ``clustering fossil". The off-diagonal correlation can be understood as the local inhomogeneities introduced by the long mode. The coupling coefficient between short modes can be modeled from the squeezed-limit bispectrum and be used to pick up the specific nonlinear correlation that we use to estimate the long modes.

Throughout this paper, we use the tree-level bispectrum in standard perturbation theory to write down the quadratic estimator for the long mode in \refeq{estimator}, i.e., we only consider the coupling between the second-order short modes and the linear-order long mode. We have tested the same estimator in both second-order GridSPT and 2LPT simulations. By calculating the ensemble mean of the cross power spectrum $P_{rm}(k)$, we show that the estimator is unbiased in second-order GridSPT but biased in 2LPT. We manage to remove the bias of the recovered long mode in 2LPT simulations with the fossil kernel term measured from the 2LPT bispectrum. The results imply that the coupling coefficient from an accurate squeezed-limit bispectrum model guarantees an unbiased estimator. 

We also aim to reconstruct the long-mode power spectrum. The auto power spectrum of the recovered mode $P_{rr}$ involves the four-point correlation function in Fourier space. $P_{rr}$ contains both the long mode power spectrum and the noise power spectrum. The noise power term is minimized if we use the inverse variance weight with the assumption that the short density modes are Gaussian, which is not true in the real universe. In fact, the noise power spectrum also receives the contribution from the connected four-point correlation function, which becomes more significant as we increase the largest wavenumber of short modes for reconstruction. Therefore, it is crucial to include the connected four-point correlation function into the noise power spectrum to accurately reconstruct the long-mode power spectrum.

We also demonstrate that the cross-correlation coefficient between the recovered and true long mode is determined by the ratio between the long-mode power spectrum and noise. To get a higher cross-correlation coefficient, we can suppress the noise power spectrum by including more short modes for reconstruction. With the estimation from fourth-order GridSPT, we estimate the best cross-correlation coefficient the tree-level fossil estimator could achieve in $N$-body simulation is 0.7 at redshifts $z = 0, 1, 2, 3$ while keeping the recovered long mode unbiased.

In order to apply the fossil estimator in $N$-body simulations and enhance the reconstruction power, we need to include more short modes of larger wavenumbers for the reconstruction. Therefore, we need to use higher-order coupling terms beyond tree-level in the perturbation theory into the estimator. However, the estimator can still be biased because the perturbation theory fails on the fully nonlinear scale. Fortunately, the response approach \citep{Wagner2014, Barreira2017} enables us to measure the squeezed-limit bispectrum in mild-sized $N$-body simulations accurately, which can potentially extend $q_{\rm max}$ from quasi-linear scale to fully nonlinear scale without biasing the estimator. We leave this part to the future work. 

To make this method more feasible in future galaxy surveys and 21cmIM experiments, we need to investigate the reconstruction from the biased tracer field. Using perturbative galaxy bias expansion, we can calculate the squeezed-limit galaxy-galaxy-matter bispectrum and reconstruct the matter density long mode from the galaxy short modes. We also need to include the redshift-space distortion effect and the foreground wedge in the 21cmIM \citep{darwish2021density}. Our next step, then, is to understand how galaxy bias and the observational effects impact the reconstruction power. 

There are many interesting applications of fossil estimators. By implementing the estimator in galaxy survey, the recovered matter long mode and observed galaxy long mode naturally form multi tracers and can constrain the local-type primordial non-Gaussianity during the inflation without cosmic variance \citep{Seljak_2009, McDonald_2009} (See \cite{darwish2021density} for more details.). Furthermore, we can construct the tensor-type fossil estimator to infer the primordial gravitational waves from the galaxy short modes, if the galaxy-galaxy-gravitational wave bispectrum can be modeled accurately. Based on the clustering fossil theory, this reconstruction method could become a powerful new statistical approach to probing the large-scale structure and increasing the scientific output of the future galaxy and 21cm surveys.

\appendix

\section{Fossil Estimator in the Continuous Limit}
\label{app:fossil_continuous}
To get a fast estimation of the noise power spectrum of the reconstructed modes, we can write down the fossil estimator in the continuous limit.
\bea
\delta_r(k) &=& P^N_G (\vk) \int_q \frac{\delta(\vq)\delta(\vk-\vq)f(\vq, \vk-\vq)}{P(q)P(|\vk-\vq|)}
\nonumber
\\
&=& P^N_G (k) \int_k^{q_\max} \frac{\d q }{(2\pi)^2} q^2 \int \d \mu \frac{\delta(\vq)\delta(\vk-\vq)f(\vq, \vk-\vq)}{P(q)P(|\vk-\vq|)}.
\nonumber
\\
\eea
In the continuous limit, the noise power spectrum under the Gaussian assumption is
\bea
\label{eq:PN_theory}
P^N_G (k) &=&  \left[ \int_{k}^{q_\max} \frac{\d q }{(2\pi)^2} q^2 \int \d \mu \frac { \left| f ( \vq , \vk - \vq ) \right| ^ { 2 } } {P ( q ) P ( |\vk - \vq| ) } \right]^{-1} 
\eea
Here we have used the relation between the integration and the grid-based discrete summation over the Fourier space
\bea
\frac1V \sum  = \int \frac{\d^3 q}{(2\pi)^3}.
\eea
The range of $\mu \equiv \hat{\vk} \cdot \hat{\vq}$ is constrained by $q \ge |\vk - \vq| > k$, which is 
\bea
\label{eq:murange}
\frac{k}{2q} \le \mu \le {\rm min}\{ 1,\, \frac{q}{2k} \}.
\eea

\acknowledgments{
We want to thank Caryl Gronwall and Michael Sigel for useful comments on the early draft of this paper. ZW wants to acknowledge Hong-ming Zhu and Oliver Philcox for useful discussions and Yuanheng Wang for helpful support during the completion of this paper. This research made use of the Roar Collab Supercomputer at Penn State University. ZW and DJ acknowledge support from the National Science Foundation Grant No. AST-2307026 at Penn State University. DJ is supported by KIAS Individual Grant PG088301 at Korea Institute for Advanced Study and was supported by NASA ATP program (80NSSC18K1103) at Penn State University.
}

\bibliographystyle{JHEP}
\bibliography{main}

\end{document}